\documentclass[twocolumn,english,aps,prb,twocolum,superscriptaddress,bibnotes,amsmath,amssymb,floatfix,groupedaddress,footinbib]{revtex4-2}

\usepackage{amsmath,amsfonts,amssymb}
\usepackage[english]{babel}
\usepackage[markup=blue authormarkupposition=left]{changes}
\usepackage{color}
\usepackage[T1]{fontenc}
\usepackage[colorlinks=true,citecolor=blue,linkcolor=magenta]{hyperref}
\usepackage[utf8]{inputenc}
\usepackage{epstopdf}
\usepackage{graphicx}
\usepackage{mhchem}
\usepackage{physics}
\usepackage{siunitx}
\usepackage{soul}
\usepackage{upgreek}
\usepackage{url}

\DeclareMathOperator{\arccosh}{arccosh}

\graphicspath{{./Figures/}}

\newcommand{\EPFL}{Institute of Physics, Swiss Federal Institute of Technology Lausanne (EPFL), CH-1015 Lausanne, Switzerland}
\newcommand{\QSE}{Center for Quantum Science and Engineering, Swiss Federal Institute of Technology Lausanne (EPFL), CH-1015 Lausanne, Switzerland}
\newcommand{\MPINAT}{Department of Ultrafast Dynamics, Max Planck Institute for Multidisciplinary Sciences, D-37077 G\"{o}ttingen, Germany}
\newcommand{\GOE}{Georg-August-Universit\"{a}t G\"{o}ttingen, D-37077 G\"{o}ttingen, Germany}

\begin{document}

\title{Free-electron interaction with nonlinear optical states in microresonators}

\author{	Yujia Yang$^{1,2,\ast,\dag}$,
		Jan-Wilke Henke$^{3,4,\ast}$, 
		Arslan S. Raja$^{1,2,\ast}$, 
		F. Jasmin Kappert$^{3,4,\ast}$, 
		Guanhao Huang$^{1,2}$, 
		Germaine Arend$^{3,4}$, 
		Zheru Qiu$^{1,2}$,
		Armin Feist$^{3,4}$,
		Rui Ning Wang$^{1,2}$,
		Aleksandr Tusnin$^{1,2}$,
		Alexey Tikan$^{1,2}$,
		Claus Ropers$^{3,4,\ddag}$,
		and Tobias J. Kippenberg$^{1,2,\S}$}
\affiliation{
$^1$\EPFL\\
$^2$\QSE\\
$^3$\MPINAT\\
$^4$\GOE
}

\maketitle

%%%%%%%%%%%%%%%%%%%%%%%%%%%%%%%%%%%%%%%%%%%%%%%%%%%%%%%%%%%%%%%%%%%%%%
%%%%%%%%%%%%%%%%%%%%%%%%%%%%% Abstract %%%%%%%%%%%%%%%%%%%%%%%%%%%%%%%
%%%%%%%%%%%%%%%%%%%%%%%%%%%%%%%%%%%%%%%%%%%%%%%%%%%%%%%%%%%%%%%%%%%%%%
\noindent\textbf{
The short de Broglie wavelength and strong interaction empower free electrons to probe scattering and excitations in materials \cite{ruskaDevelopmentElectronMicroscope1987}, and resolve the structure of biomolecules \cite{dubochetDevelopmentElectronCryoMicroscopy2018}.
Recent advances in using nanophotonic structures to mediate bilinear electron-photon interaction have brought novel optical manipulation schemes to electron beams.
This has enabled high space-time-energy resolution electron microscopy \cite{barwickPhotoninducedNearfieldElectron2009a, ryabovElectronMicroscopyElectromagnetic2016b, polmanElectronbeamSpectroscopyNanophotonics2019a, wangCoherentInteractionFree2020e, kfirControllingFreeElectrons2020c} quantum-coherent optical modulation \cite{feistQuantumCoherentOptical2015b, henkeIntegratedPhotonicsEnables2021}, attosecond metrology and pulse generation \cite{priebeAttosecondElectronPulse2017, morimotoDiffractionMicroscopyAttosecond2018, ryabovAttosecondMetrologyContinuousbeam2020b}, transverse electron wavefront shaping \cite{garciadeabajoElectronDiffractionPlasmon2016, vanacoreUltrafastGenerationControl2019b, konecnaElectronBeamAberration2020, feistHighpurityFreeelectronMomentum2020a, madanUltrafastTransverseModulation2022}, dielectric laser acceleration \cite{englandDielectricLaserAccelerators2014a, sapraOnchipIntegratedLaserdriven2020}, and electron-photon pair generation \cite{feistCavitymediatedElectronphotonPairs2022a}. 
However, photonic nanostructures also exhibit nonlinearities, which have to date not been exploited for electron-photon interactions.
Here, we report the interaction of electrons with spontaneously generated Kerr nonlinear optical states inside a continuous-wave driven photonic chip-based microresonator.
Optical parametric processes give rise to spatiotemporal pattern formation, or `dissipative structures', corresponding to coherent or incoherent optical frequency combs \cite{kippenbergDissipativeKerrSolitons2018}.
By coupling such `microcombs' \textit{in situ} to electron beams, we demonstrate that different dissipative structures induce distinct fingerprints in the electron spectra and in Ramsey-type interference patterns of secant trajectories. 
In particular, using spontaneously formed femtosecond temporal solitons, we achieve ultrafast temporal gating of the electron beam without the necessity of a pulsed laser source or a pulsed electron source. 
Our work elucidates the interaction of free electrons with a variety of nonlinear dissipative states, demonstrates the ability to access solitons inside an electron microscope, and extends the use of microcombs to unexplored territories, with ramifications in novel ultrafast electron microscopy, light-matter interactions driven by on-chip temporal solitons, and ultra-high spatiotemporal resolution sampling of nonlinear optical dynamics and devices.
}

%%%%%%%%%%%%%%%%%%%%%%%%%%%%%%%%%%%%%%%%%%%%%%%%%%%%%%%%%%%%%%%%%%%%%%
%%%%%%%%%%%%%%%%%%%%%%%%%%%%%%%%%%%%%%%%%%%%%%%%%%%%%%%%%%%%%%%%%%%%%%
%%%%%%%%%%%%%%%%%%%%%%%%%%%%%%%%%%%%%%%%%%%%%%%%%%%%%%%%%%%%%%%%%%%%%%

%%%%%%%%%%%%%%%%%% Fig. 1 %%%%%%%%%%%%%%%%%%
\begin{figure*}[ht]
\centering
\includegraphics[width=\textwidth]{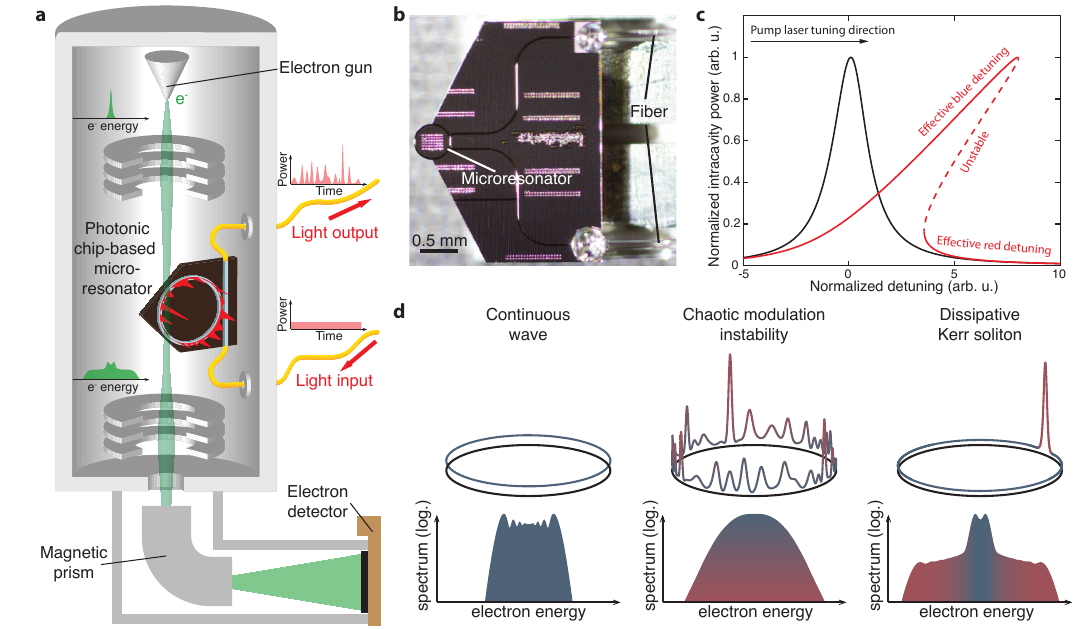}
\caption{
\textbf{Free-electron interaction with nonlinear optical states in a photonic chip-based optical microresonator.}
\textbf{a}, Schematic of the experiment. Electrons in a transmission electron microscope (TEM) pass by a fiber-coupled photonic chip-based \ce{Si3N4} microresonator with anomalous dispersion and Kerr nonlinearity. The electron beam direction is parallel to the chip surface (aloof configuration). Intracavity four-wave mixing leads to optical frequency comb generation and a related intracavity field amplitude modulation. Optical fibers attached to the chip permit control over the optical input and access to the optical output. Stimulated inelastic scattering between electrons and optical waveforms induces electron spectral broadening, which is measured with an electron spectrometer equipped with an event-based electron detector. 
\textbf{b}, Photo of the fiber-packaged \ce{Si3N4} photonic chip mounted on a custom TEM sample holder.
\textbf{c}, Normalized intracavity power versus laser detuning showing a resonance tilt and bistability due to the Kerr effect.
\textbf{d}, Illustration of intracavity waveforms and post-interaction electron spectra for continuous wave, chaotic modulation instability, and dissipative Kerr soliton states. The electron spectral broadening originates from incoherent summation of electrons which have interacted with the intracavity field at different times.
}
\label{Fig:1}
\end{figure*}
%%%%%%%%%%%%%%%%%% Fig. 1 %%%%%%%%%%%%%%%%%%

Nonlinear optical phenomena are widely used in science and technology alike; they allow broadband coherent supercontinuum generation that has unlocked optical frequency metrology \cite{dudleySupercontinuumGenerationPhotonic2006}, squeezed light generation as used for advanced gravitational wave astronomy \cite{groteFirstLongTermApplication2013},  optical parametric oscillation for lasers in hard-to-access wavelength regimes \cite{myersQuasiphasematchedOpticalParametric1995}, and entangled photon pair generation \cite{kwiatNewHighIntensitySource1995} that is pivotal to quantum information science, to name only a few examples. 
Over the past decade, triggered by advances in ultra-low loss (i.e. ultra-high quality-factor $Q$) photonic microresonators, continuous-wave driven microresonators with Kerr nonlinearity (i.e. $\chi^{(3)}$) have been shown to give rise to a host of `dissipative structures' \cite{kippenbergDissipativeKerrSolitons2018}.
In particular, dissipative Kerr solitons \cite{herrTemporalSolitonsOptical2014c} in the anomalous dispersion (and their `platicon' counterparts in the normal dispersion) lead to coherent optical frequency combs with wide-ranging applications from atomic clocks \cite{pappMicroresonatorFrequencyComb2014}, terabit communications \cite{marin-palomoMicroresonatorbasedSolitonsMassively2017} to photonic computing \cite{feldmannParallelConvolutionalProcessing2021a} and astrophysical spectroscopy \cite{obrzudMicrophotonicAstrocomb2019a, suhSearchingExoplanetsUsing2019}.
Recently, high-$Q$ silicon nitride (\ce{Si3N4}) photonic chip-based microresonators have been used to demonstrate cavity-enhanced electron-light scattering \cite{henkeIntegratedPhotonicsEnables2021} and vacuum field-induced spontaneous parametric generation of electron-photon pairs \cite{feistCavitymediatedElectronphotonPairs2022a}. 
Despite leading to highly nonlinear free-electron transitions that involve the exchange of many quanta of light, to date, only the linear cavity response was exploited to resonantly enhance the intracavity field and the electron-light coupling strength.
Nonlinear optical responses have been predicted to endow electron-light interaction with new capabilities \cite{konecnaNanoscaleNonlinearSpectroscopy2020b, coxNonlinearInteractionsFree2020, garciadeabajoCompleteExcitationDiscrete2022}, but have not been experimentally demonstrated so far.
Kerr microresonators support diverse intracavity nonlinear dissipative structures, including dissipative Kerr solitons \cite{herrTemporalSolitonsOptical2014c}, Turing patterns \cite{huangGloballyStableMicroresonator2017}, chaotic modulation instabilities \cite{delhayeOpticalFrequencyComb2007a, hanssonDynamicsModulationalInstability2013}, breathing solitons \cite{yuBreatherSolitonDynamics2017, lucasBreathingDissipativeSolitons2017a}, soliton crystals \cite{coleSolitonCrystalsKerr2017a, karpovDynamicsSolitonCrystals2019b}, and switching waves \cite{andersonZeroDispersionKerr2022}.   
Here, we study the coupling of electron beams with such spatiotemporal patterns in a photonic chip-based Kerr microresonator, extending cavity-mediated electron-light interactions to the nonlinear optical regime.

%%%%%%%%%%%%%%%%%% Fig. 2 %%%%%%%%%%%%%%%%%%
\begin{figure*}[ht]
\centering
\includegraphics[width=\textwidth]{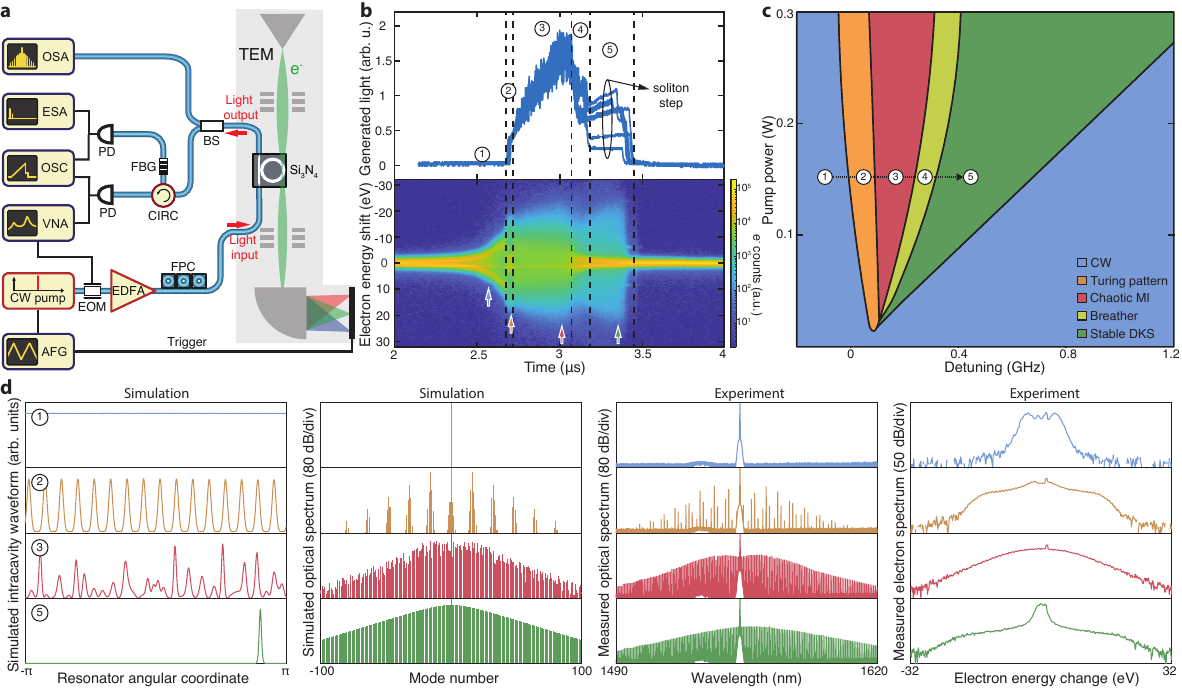}
\caption{
\textbf{Electron spectral fingerprints imprinted by microresonator-based nonlinear dissipative structures}. 
\textbf{a}, Experimental setup. OSA: optical spectrum analyzer; ESA: electronic spectrum analyzer; OSC: oscilloscope; VNA: vector network analyzer; AFG: arbitrary waveform generator; PD: photodetector; EOM: electro-optic modulator; EDFA: erbium-doped fiber amplifier; FBG: fiber Bragg grating; CIRC: optical circulator; FPC: fiber polarization controller; BS: beam splitter. 
\textbf{b}, Oscilloscope traces of the generated light (upper panel) and the detuning-dependent electron spectrum (lower panel) when scanning the pump laser frequency from the blue to the red side of the resonance. The electron spectra are summed over multiple scans. The dashed lines and the numbers indicate several regions corresponding to different intracavity optical states. 
\textbf{c}, Simulated stability chart of the \ce{Si3N4} microresonator showing the two dimensional parameter space of optical pump power and frequency detuning. Each colored region represents the sub-space for the existence of the labeled state. The dashed arrow depicts the experimental trajectory of the detuning scan shown in \textbf{b}.
\textbf{d}, Simulated intracavity optical waveforms and optical spectra, as well as measured optical and electron spectra for continuous wave, Turing pattern, chaotic modulation instability (MI), and single dissipative Kerr soliton (DKS) states (top to bottom). The numbering corresponds to that in \textbf{b}\&\textbf{c}, and the color encoding is equivalent to \textbf{c}. The measured electron spectra are vertical slices of the detuning-dependent electron spectrum in \textbf{b}, with the detuning indicated by the arrows of corresponding colors.
}
\label{Fig:2}
\end{figure*}
%%%%%%%%%%%%%%%%%% Fig. 2 %%%%%%%%%%%%%%%%%%

\section{Transient observation of electron spectral modulation by nonlinear optical states}
We establish a new approach to studying and harnessing nonlinear optical effects in a transmission electron microscope (TEM), building on the recently developed platform based on ultra-low loss \ce{Si3N4} photonic chip-based microresonators \cite{henkeIntegratedPhotonicsEnables2021, feistCavitymediatedElectronphotonPairs2022a}. 
In this scheme, highlighted in Fig. \ref{Fig:1}a, a photonic chip-based microresonator is placed in the TEM with the quasi-monochromatic electron beam (e-beam) passing over the resonator surface in an aloof geometry.
The fiber-packaged \ce{Si3N4} photonic chip-based microresonator (Fig. \ref{Fig:1}b) exhibits a high quality-factor of $\sim3.0\times10^6$ and anomalous group velocity dispersion to resonantly enhance the intracavity field and facilitate nonlinear frequency mixing. 
The experiments are repeatedly performed on multiple microresonators.
Here, we show the data acquired from three chips (denoted by letters A, B, and C; see Methods for wafer identification).
When the microresonator is pumped with continuous-wave (CW) input light above threshold \cite{kippenbergKerrNonlinearityOpticalParametric2004}, the Kerr nonlinearity causes a resonance tilt and cavity bistability when scanning the pump frequency from the blue to the red side of the resonance (Fig. \ref{Fig:1}c), eventually leading to the generation of an incoherent or coherent optical frequency comb via cascaded four-wave mixing (FWM) \cite{kippenbergDissipativeKerrSolitons2018}. 
In the time domain, the optical spectra correspond to the spontaneous formation of diverse nonlinear dissipative structures (i.e. spatiotemporal patterns), with a couple of examples shown in Fig. \ref{Fig:1}d. 

Electrons traversing the optical evanescent near field of the air-cladded microresonator undergo inelastic electron-light scattering (IELS), leading to the formation of `photon sidebands' in the electron spectrum spaced by integer multiples of the photon energy, corresponding to the absorption and emission of photons by the electron \cite{barwickPhotoninducedNearfieldElectron2009a}.
The intensity of these sidebands as well as the overall width of the spectrum are determined by a coupling parameter $g$ \cite{parkPhotoninducedNearfieldElectron2010}, which is sensitive to the electric field component parallel to the electron momentum over the interaction time:
\begin{equation}\label{eq:g_def_main}
     g(x,y) = - \frac{e}{2 \hbar \omega_0} \int_{-\infty}^{+\infty} E_z(x,y,z') e^{- i\frac{\omega_0}{v}z'} \,dz',
\end{equation}
where $z$ is the coordinate along the e-beam direction, $(x,y)$ the transverse coordinate, $e$ the electron charge, $\hbar$ the reduced Planck constant, $\omega_0$ the angular frequency of the optical carrier wave, and $v$ the electron velocity.
In a simplified picture, a continuous electron beam therefore samples the electric field of the intracavity waveforms, at random arrival times of the electrons within the beam.
This results in a characteristic electron spectral shape for each of the nonlinear optical intracavity states (Fig. \ref{Fig:1}d).
In the experiment, such electron spectra are acquired by an electron spectrometer, while the concomitant optical measurements are performed on the outcoupled light.

The experimental setup (Fig. \ref{Fig:2}a) consists of a TEM with an imaging spectrometer, a chip-based nonlinear microresonator, and a setup for optical excitation as well as spectral and temporal characterization of the output light.
We use the transverse electric mode family and \SI{200}{\keV} electrons throughout the experiment. 
We transiently generate nonlinear optical states in Chip A by continuously scanning the pump laser frequency across a cavity resonance, and record electron energy spectra in parallel. 
Figure \ref{Fig:2}b shows several oscilloscope traces of the generated light obtained by rejecting the pump laser from the output light, together with the detuning-dependent electron spectra (only blue-to-red detuning scan shown).
The oscilloscope traces bear the characteristics of a detuning scan for a Kerr nonlinear cavity with anomalous dispersion, exhibiting a typical `soliton step' \cite{herrTemporalSolitonsOptical2014c} that demonstrates the ability to generate a DKS inside a TEM.
We identify five regions corresponding to distinct intracavity states: (1) at low pump power, the nonlinear optical response is negligible, and the intracavity field is a monochromatic CW optical field; (2) with an increased detuning, and thus a higher intracavity power, cascaded FWM leads to the generation of new frequency components and an amplitude modulation of the intracavity field, known as a Turing pattern or cnoidal wave; (3) when further red-detuning the pump frequency, the intracavity optical state enters the chaotic modulation instability (MI) regime, forming a disordered and rapidly varying waveform and generating an incoherent Kerr frequency comb; (4) when the pump is red-detuned from the cavity resonance (i.e. landing on the `soliton step'), localized structures are spontaneously formed and self-stabilized, which at small detunings consist of breathing solitons with periodically oscillating temporal and spectral shapes; (5) with an increased detuning on the soliton step, shape-invariant stable DKS states are generated, featuring one or multiple femtosecond temporal pulses forming a coherent frequency comb.
The existence of these states is determined by two key parameters: the optical pump power and the detuning of the pump frequency to the cavity resonance.
The simulated stability chart in terms of these two parameters (Fig. \ref{Fig:2}c) agrees well with the experimental measurement, illustrating the generation of different nonlinear intracavity states while scanning the pump detuning (dashed arrow).
 
We next consider how the electron beam interacts with these intracavity states.
With the polar-to-Cartesian coordinate transformation, the IELS coupling parameter (Eq. \ref{eq:g_def_main}) becomes
\begin{equation} \label{eq:g_cav_main}
    \begin{aligned}
             g(x,y) = - \frac{e}{2 \hbar \omega_0} &\int_{-\infty}^{+\infty} A(\phi+\phi_0+D_1\frac{z'}{v},\tau)(-e_\phi(x,y)\cos\phi\\
             &-e_r(x,y)\sin\phi)e^{-in\frac{\omega_0}{c}R\phi}e^{- i\frac{\omega_0}{v}z'} \,dz',
    \end{aligned}
\end{equation}
where $A(\phi,\tau)$ is the intracavity waveform (in the frame rotating at the optical group velocity) with the microresonator angular coordinate $\phi$ and the slow time $\tau$ \cite{kippenbergDissipativeKerrSolitons2018} (much larger than the cavity round-trip time), $D_1/2\pi$ is the microresonator free spectral range (FSR), $e_{\phi}$ and $e_r$ are the tangential and radial optical modal fields, respectively, $n$ is the linear refractive index, $c$ is the speed of light, $R$ is the microresonator ring radius, and $\phi_0$ accounts for the angular offset of the rotating frame from the laboratory frame. 
The coupling parameter further determines the electron spectrum via the $N$-th photon sideband amplitude $J_N(2|g(x,y)|) (\frac{g(x,y)}{|g(x,y)|})^N$, where $J_N$ is the Bessel function of the first kind.
The spectra of a continuous electron beam are obtained by an incoherent averaging over the angular offset $\phi_0$ and the slow time $\tau$ (see Methods and Supplementary Note 1).

Figure \ref{Fig:2}d shows the simulated intracavity waveforms and optical spectra as well as the measured optical and electron spectra for four intracavity states.
The electron spectra in Figs. \ref{Fig:2}b\&d demonstrate distinct characteristic shapes for different states. 
For the CW state, the electron spectrum has the typical double-peak shape from coherent phase modulation, and an approximate spectral width of $4|g|\hbar\omega_0$ \cite{feistQuantumCoherentOptical2015b, giulioProbingQuantumOptical2019a, henkeIntegratedPhotonicsEnables2021} (Fig. \ref{Fig:2}b blue arrow, Fig. \ref{Fig:2}d first row). 
For an increasing pump detuning, the intracavity state undergoes a transition to a Turing pattern and chaotic MI. 
The electron spectra broaden due to an increasing average intracavity power, together with the appearance of a Gaussian-shaped background and a reduction of the spectral double peaks (Fig. \ref{Fig:2}b yellow and red arrows, Fig. \ref{Fig:2}d second and third rows).
These observations can be attributed to the intracavity field strength no longer being uniform. 
When further increasing the detuning, stable DKS (multiple or single soliton) states are stochastically generated, and the corresponding electron spectrum features a strong, narrow central peak on a weak, broad plateau (Fig. \ref{Fig:2}b green arrow, Fig. \ref{Fig:2}d fourth row).
Our results show that nonlinear intracavity states lead to unique fingerprints in the electron spectra.

%%%%%%%%%%%%%%%%%% Fig. 3 %%%%%%%%%%%%%%%%%%
\begin{figure*}[ht]
\centering
\includegraphics[width=\textwidth]{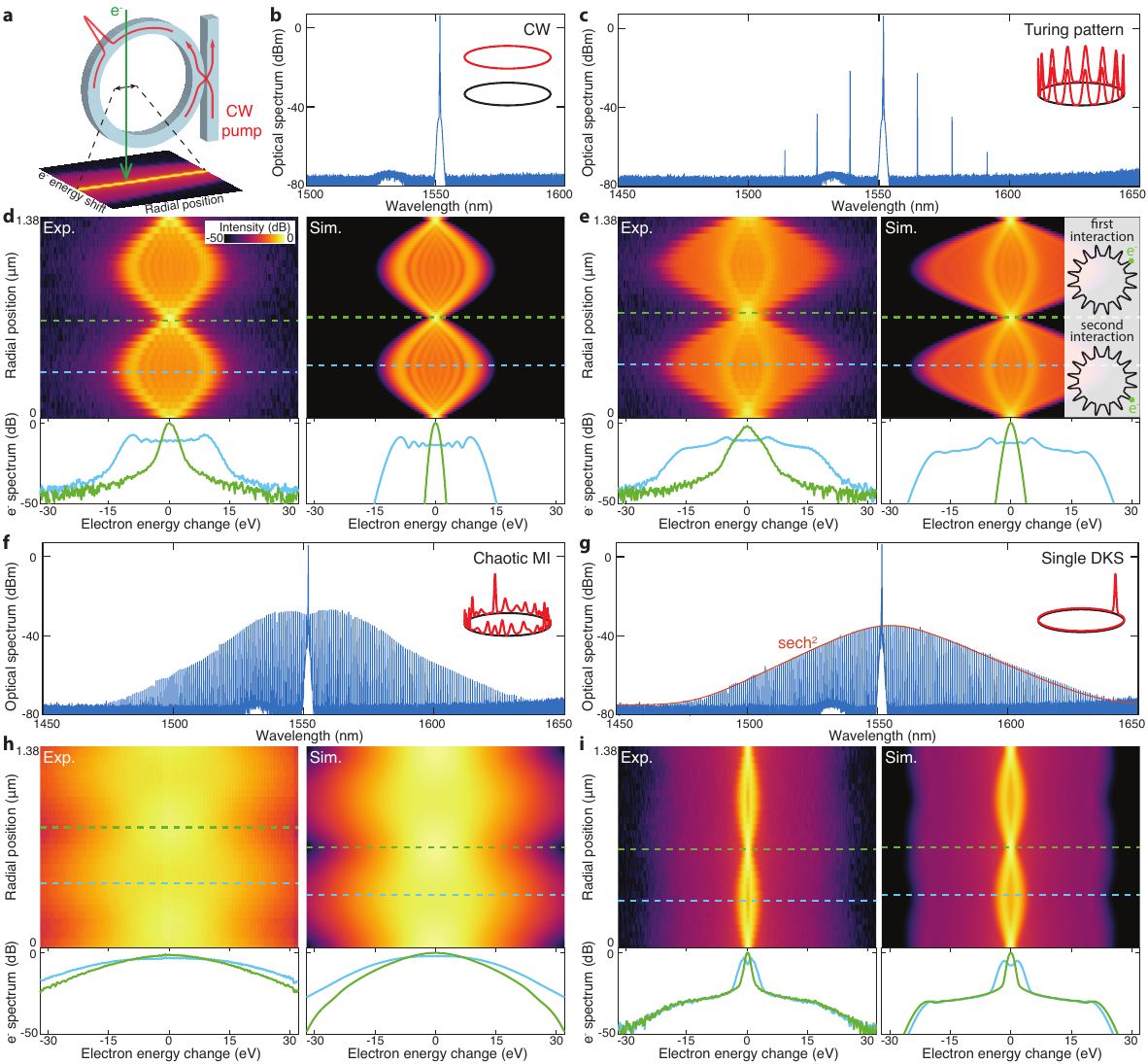}
\caption{
\textbf{Ramsey-type interference of nonlinear intracavity states}. 
\textbf{a}, Illustration of the Ramsey-type interferometry configuration in which the electron trajectory intersects twice with the microresonator waveguide. A single dissipative Kerr soliton (DKS) consisting of a temporal soliton pulse and a weak CW background is depicted as an exemplary intracavity state. The phase of the Ramsey interference is tuned by changing the electron beam position and hence the time delay between the two interactions. The color image at the bottom shows exemplary position-dependent electron spectra.
\textbf{b}, \textbf{c}, \textbf{f}, \textbf{g}, Measured optical spectra for CW (\textbf{b}), Turing pattern (\textbf{c}), chaotic modulation instability (MI) (\textbf{f}), and single DKS (\textbf{g}) states. 
\textbf{d}, \textbf{e}, \textbf{h}, \textbf{i}, Measured (left panel, Exp.) and simulated (right panel, Sim.) Ramsey interference electron spectral patterns for the 4 states corresponding to the optical spectra. 
The range of radial positions for all interference patterns is \SI{1.375}{\um}. For each interference pattern, electron spectra of two line cuts (green and blue dashed lines) are plotted at the bottom. The schematic in \textbf{e} depicts snapshots of the Turing pattern intracavity waveforms at the time of the two interactions for one exemplary electron arrival time (green dots). The $\mathrm{sech^2}$ fitting of the optical spectrum in \textbf{g} corresponds to a DKS pulse duration of $\sim\SI{98.5}{\fs}$.
}
\label{Fig:3}
\end{figure*}
%%%%%%%%%%%%%%%%%% Fig. 3 %%%%%%%%%%%%%%%%%%

%%%%%%%%%%%%%%%%%% Fig. 4 %%%%%%%%%%%%%%%%%%
\begin{figure*}[ht]
\centering
\includegraphics[width=\textwidth]{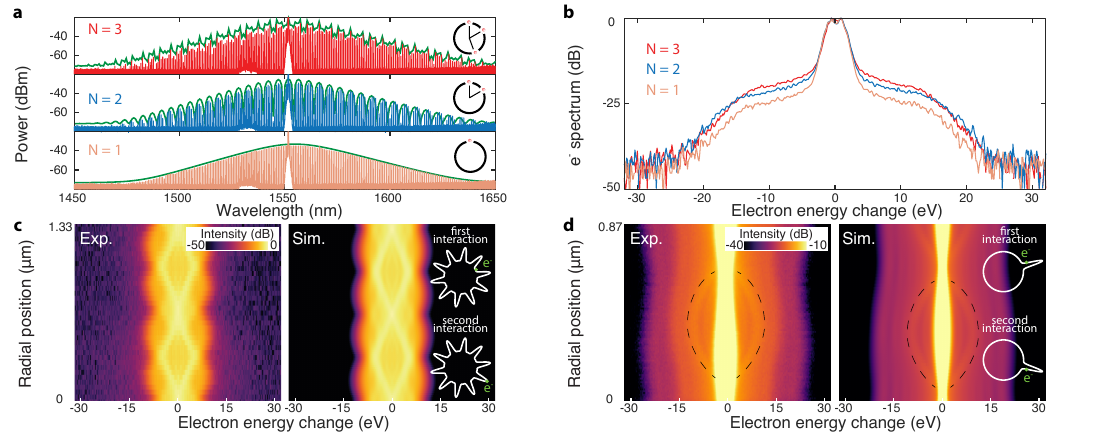}
\caption{
\textbf{Electron modulation from coherent intensity-varying dissipative structures}. 
\textbf{a}, Optical spectra of multi-soliton states generated via backward switching. The insets show the number and relative phase of soliton pulses retrieved by fitting the spectra (green curves).   
\textbf{b}, Corresponding electron spectra from the modulation by multi-soliton states.
\textbf{c\&d}, Experimental (left panel, Exp.) and simulated (right panel, Sim.) Ramsey interference patterns for a Turing pattern in Chip B (\textbf{c}) and a single-soliton state in Chip C (\textbf{d}). Insets in \textbf{c\&d} illustrate the intracavity optical waveforms (white) coupled to the electron (green dot) in the two interaction regions. Black dashed lines in \textbf{d} outline the halo feature observed in both the experimental and simulated Ramsey patterns, which results from the double interactions of the electron with the soliton.
}
\label{Fig:5}
\end{figure*}
%%%%%%%%%%%%%%%%%% Fig. 4 %%%%%%%%%%%%%%%%%%

\section{Fingerprints of dissipative structures in Ramsey-type interference}
We gain further insight into the interaction of free electrons with nonlinear dissipative structures by investigating the Ramsey-type interferences of two sequential interactions with the microresonator.
As depicted in Fig. \ref{Fig:3}a, the e-beam intersects the waveguide near field twice, resulting in constructive or destructive interference due to a position-dependent phase delay between the two interactions \cite{echternkampRamseytypePhaseControl2016a, henkeIntegratedPhotonicsEnables2021}.
For dissipative structures, the evolution of both the optical carrier and envelope between the two regions provides new degrees of freedom in free-electron modulation well beyond that of coherent phase modulation at a uniform field strength.
As an example, Fig. \ref{Fig:3}a illustrates electron spectra observed for various e-beam positions along the chip surface while modulating the electrons by a single DKS state consisting of a temporal soliton pulse and a weak CW background.
We change the Ramsey interference phase delay by scanning the e-beam position along the surface of Chip A (i.e. changing the radial position of the e-beam with respect to the ring center), and acquire position-dependent electron spectra while maintaining a given intracavity state (Fig. \ref{Fig:3}). 

For the monochromatic CW field (Fig. \ref{Fig:3}b), the electron spectral width oscillation along the radial position has the typical pattern of previously reported Ramsey-type interferences \cite{henkeIntegratedPhotonicsEnables2021} (Fig. \ref{Fig:3}d). The nodes (minimal spectral width) of the oscillation results from destructive interference arising from a $\pi$ phase shift of the optical field at the two interactions.

For a Turing pattern, the generated frequency components and the accompanying amplitude modulation (Fig. \ref{Fig:3}c) alter the interaction-induced spectrum of the electrons. 
The Ramsey interference pattern in Fig. \ref{Fig:3}e has a central, low-energy-change part of the spectrum resembling the double-peak shape from a CW interaction, but with a narrower spectral width.
This is caused by a reduced pump-line power due to intracavity FWM, despite a higher average power.
Additionally, the electron spectrum has a broad shoulder that is absent in the case of CW field.
This shoulder arises from the interaction of electrons with the intensity peaks of the intracavity waveform, as illustrated in the inset of Fig. \ref{Fig:3}e.  
Note that for the chosen e-beam position, the temporal periodicity of the Turing pattern amplitude modulation is approximately commensurate with the travel time difference of the electron and the optical envelope between the two interaction regions ($\Delta T_{\mathrm{travel}}\approx NT_{\mathrm{Turing}}$, $N\in\mathbb{Z}$).
Specifically, the electron experiences nearly equal optical intensities in the two interactions due to the time-translation invariance of the periodic Turing pattern; this leads to similar modulation strengths of the two interactions just like the CW case.
Hence, the low-energy-change part and the broad shoulder jointly increase and decrease in their spectral widths.

%%%%%%%%%%%%%%%%%% Fig. 5 %%%%%%%%%%%%%%%%%%
\begin{figure*}[ht]
\centering
\includegraphics[width=\textwidth]{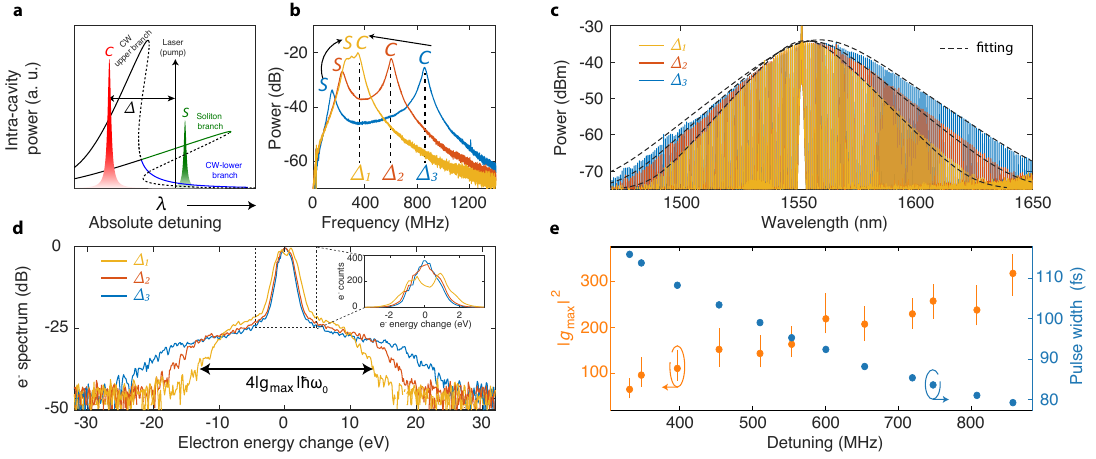}
\caption{
\textbf{Electron probing of single dissipative Kerr solitons}. 
\textbf{a}, Resonance bistability in the presence of the Kerr effect (self-phase modulation). A soliton is formed when the high power upper continuous-wave (CW) branch (black) decays to the soliton branch (green). In the soliton state, the pump is effectively red-detuned, and both soliton branch and CW low-power branch (blue) exist.   
\textbf{b}, Double resonance response of soliton ($\mathcal{S}$) and cavity ($\mathcal{C}$) observed by scanning electro-optic-modulation sidebands of the pump via a vector network analyzer (VNA). This allows to determine the effective detuning, which is directly linked to the spectral width of soliton spectrum (pulse duration).
\textbf{c}, The soliton spectra acquired at three different effective detunings ($\Delta$) as indicated in \textbf{b} ($\Delta_1$,$\Delta_2$,$\Delta_3$) and the corresponding electron spectra in \textbf{d}. 
\textbf{e}, The dependence of the dimensionless coupling parameter |$g_{\mathrm{max}}$|$^2$ and soliton pulse duration on the effective pump detuning. The pulse width is estimated by fitting the soliton spectrum, while |$g_{\mathrm{max}}$|$^2$ is obtained by fitting the electron spectrum. 
}
\label{Fig:4}
\end{figure*}
%%%%%%%%%%%%%%%%%% Fig. 5 %%%%%%%%%%%%%%%%%%

For the chaotic MI corresponding to an incoherent Kerr comb and a disordered waveform (Fig. \ref{Fig:3}f), the characteristic double peaks from uniform phase modulation can no longer be identified in the electron spectrum, which instead has a smooth Gaussian shape.
We further find a reduced spatial dependence and absence of nodes in the position-dependent scan (Fig. \ref{Fig:3}h), arising from random intracavity field strengths in this state that prohibit well-defined interference conditions of the two interactions.
Notably, the e-beam modulated by the chaotic MI state possesses a spectrum similar to that in a previous report on electron interaction with thermal states \cite{dahanImprintingQuantumStatistics}. 
Rather than invoking photon statistics, our analysis attributes the spectral shape to the statistical fluctuations of the intracavity optical intensity, which is stochastically sampled by the electrons and averaged on the detector (Supplementary Note 5).
 
In the stable DKS state, a single temporal soliton pulse spontaneously forms and circulates in the cavity on a weak CW background (cf. Fig. \ref{Fig:2}d and Fig. \ref{Fig:3}a).
The waveform can be written analytically as \cite{herrTemporalSolitonsOptical2014c} 
\begin{equation}\label{eq:DKS}
     \Psi \simeq \Psi_0 + \sqrt{\frac{4\Delta}{\kappa}} e^{i\theta_0} \sech(\sqrt{\frac{2\Delta}{D_2}}\phi),
\end{equation}
where $\Delta$ denotes the pump detuning, $\kappa$ the cavity decay rate, $\theta_0$ the soliton phase, and $D_2$ the second order dispersion. 
In the frequency domain, a coherent, low-noise optical frequency comb with a characteristic sech$^2$ spectral shape is generated (Fig. \ref{Fig:3}g). 
We observe that the corresponding electron spectra have a unique shape featuring a strong, narrow low-energy-change region, and a weak, broad plateau (Fig. \ref{Fig:3}i). 
The former exhibits the hallmark of a coherent phase modulation produced by the coupling of the electrons to the weak CW background. 
The latter, on the other hand, has a low total spectral weight and a broad width originating from the interaction between electrons and the short-duration, high-peak-power soliton pulse. 
As the DKS pulse duration ($\sim\SI{100}{fs}$) is much shorter than the round-trip time ($\sim\SI{10}{\ps}$), only a small fraction of the electrons in the continuous e-beam interact with the DKS pulse, and hence the plateau is much lower than the central low-energy-change region.
In addition, the plateau has a moderate, position-dependent spectral width oscillation, arising from the Ramsey interference of electrons interacting with the soliton pulse and the CW background one time each.
This oscillation has the same period as that of the CW Ramsey pattern, but with an offset which, notably, encodes the soliton phase (Supplementary Note 4).

For all four quintessential intracavity states, the experimental electron spectra and Ramsey patterns are well-reproduced by the simulations (Fig. \ref{Fig:3}). 
The theoretical consideration combines approximate solutions of the reduced time-dependent Schr\"{o}dinger equation for the electron, and the Lugiato-Lefever equation for the light, amounting to the nonlinear Schr\"{o}dinger equation for optics in a damped, driven, nonlinear Kerr microresonator (see Methods and Supplementary Note 1).  

The diverse intracavity optical intensity variations generate numerous additional features in electron spectra, with three examples highlighted in Fig. \ref{Fig:5}.
First, the Kerr microresonator also supports multi-soliton states with multiple DKS pulses circulating in the cavity. 
Experimentally, we generate different multi-soliton states in Chip A via the backward switching technique \cite{guoUniversalDynamicsDeterministic2017}, and record the optical and electron spectra associated with three, two, and one-soliton states (Fig. \ref{Fig:5}a\&b). 
The relative number of electrons in the spectral plateaus is 2.82:1.87:1 for the three, two, and single-soliton state, respectively.

In addition, the Ramsey interference configuration gives rise to additional degrees of freedom for electron spectral modulation.
Figure \ref{Fig:5}c shows the experimental and simulated Ramsey interference patterns for a Turing pattern in Chip B.
Note that the Ramsey pattern is drastically different from the one in Fig. \ref{Fig:3}e, since here the temporal periodicity of the Turing pattern amplitude modulation is incommensurate with the travel time difference of the electron and the optical envelope between the two interaction regions ($\Delta T_{\mathrm{travel}}\approx(N+1/2)T_{\mathrm{Turing}}$) as depicted in the inset.

Similarly, the Ramsey interference pattern (Fig. \ref{Fig:5}d) for a single-DKS state in Chip C possesses additional features - a hemispherical halo (indicated by the black dashed lines) - compared to the pattern in Fig. \ref{Fig:3}i. 
This halo feature is produced by electrons interacting twice with the DKS pulse - the pulse's trailing edge in the first interaction and the leading edge in the second interaction (Supplementary Note 3).

\section{Probing dissipative Kerr soliton dynamics with free electrons}
We utilize IELS to probe basic DKS properties (Fig. \ref{Fig:4}).
Kerr microresonators possess a resonance bi-stability, and in DKS generation the high power CW branch decays into the lower soliton branch while the pump is red-detuned (Fig. \ref{Fig:4}a).
An adjustment of the effective detuning $\Delta$ alters the DKS properties, including the soliton pulse duration, peak field, and background CW amplitude. 
Figure \ref{Fig:4}b shows a DKS state in Chip A probed by scanning the electro-optic-modulation sidebands of the pump via a vector network analyzer (cf. Fig. \ref{Fig:2}a) for three different pump detunings (large, medium, small). 
The effective detuning can then be obtained from the soliton ($\mathcal{S}$) and cavity ($\mathcal{C}$) resonances in these traces \cite{guoUniversalDynamicsDeterministic2017}. 
The measured optical spectra (Fig. \ref{Fig:4}c) illustrate the detuning-dependent DKS frequency comb width, with the increasing detuning leading to a broader spectrum and hence a shorter DKS pulse duration $\tau_{\mathrm{FWHM}} \simeq \frac{2\arccosh(\sqrt{2})}{D_1}\cdot \sqrt{\frac{D_2}{2 \Delta}} $.

When measuring the electron spectrum, an increasing detuning $\Delta$ leads to a wider plateau, induced by an increasing DKS peak field ($\sim\sqrt{4\Delta/\kappa}$) (Fig. \ref{Fig:4}d). 
Meanwhile, the plateau height slightly decreases, since the DKS pulse duration $\tau_{\mathrm{FWHM}}$ decreases and less electrons interact with the pulse.
A larger detuning moreover leads to a weaker CW background $\Psi_0$, which is also reflected by the weaker modulation as seen in the central low-energy-change region of the electron spectra (inset in Figs. \ref{Fig:4}d).
Figure \ref{Fig:4}e illustrates the detuning dependence of the plateau width (approximated by $4|g_{\mathrm{max}}| \hbar \omega_0$ with $g_{\mathrm{max}}$ the maximum coupling parameter) measured from electron spectra, and the DKS pulse duration extracted from optical spectra. 
As $|g_{\mathrm{max}}|$ is proportional to the DKS peak field $\sim\sqrt{4\Delta/\kappa}$, $|g_{\mathrm{max}}|^2$ scales linearly with the detuning $\Delta$.
Additional data from a continuous detuning scan can be found in Supplementary Note 7.

%%%%%%%%%%%%%%%%%%%%%%%%%%%%%%%%%%%%%%%%%%%%%%%%%%%%%%%%%%%%%%%%%%%%%%
%%%%%%%%%%%%%%%%%%%%%%%%%%%%%%%%%%%%%%%%%%%%%%%%%%%%%%%%%%%%%%%%%%%%%%
%%%%%%%%%%%%%%%%%%%%%%%%%%%%%%%%%%%%%%%%%%%%%%%%%%%%%%%%%%%%%%%%%%%%%%
 
\section{Discussion and outlook}
In conclusion, we extend free-electron-light interaction to the nonlinear optics regime harnessing optical frequency combs in a Kerr nonlinear microresonator.
Our results unlock the potential to non-invasively probe ultrafast transient nonlinear optical dynamics with nanometer-femtosecond spatiotemporal resolution and direct access to the intracavity field. 
This work also opens new avenues for optical manipulation of free electrons beyond the regimes of pulsed or continuous-wave lasers. 
The integrated photonics toolbox provides great diversity and flexibility for tailoring on-chip optical waveforms and frequency components, thus promising advanced electron control schemes via arbitrary optical waveform generation and frequency conversion, potentially enabling arbitrary electron wavefunction generation. 
Combining free-electron sampling with tailored optical waveforms leads to flexible and programmable control on the amplitudes of photon sidebands, thus providing a route towards free-electron boson sampling for quantum information processing \cite{talebiStrongInteractionSlow2020, chahshouriTailoringNearfieldmediatedPhoton2023}.

Furthermore, we achieve ultrafast electron-light interaction in an innovative scheme with chip-based femtosecond temporal solitons, in the absence of pulsed lasers or pulsed electron sources. 
This scheme combines the strong electron-light interaction enabled by the resonantly enhanced intracavity field shown previously \cite{henkeIntegratedPhotonicsEnables2021} with electron coupling to spontaneously formed, spatiotemporally confined solitons in a nonlinear microresonator.
This work will facilitate ultrafast electron microscopy in a conventional TEM equipped with a photonic chip and a CW laser, using temporal photon-gating \cite{hassanPhotonGatingFourdimensional2015, fuNanoscalefemtosecondDielectricResponse2020} with dissipative Kerr solitons instead of mode-locked lasers.
Spectral filtering of time-gated electrons could be further simplified by using `dark pulses' formed in a normal dispersion microresonator, as only the pulsed region will allow the electron to retain its original energy.
The electrons regulated by the DKS pulses form a train of pulses with sub-100-fs duration and a GHz-to-THz repetition rate, which is orders of magnitude higher than that of state-of-the-art ultrafast TEMs and allows a much higher beam current. 
The temporal resolution could be brought to the sub-10-fs regime with advances in photonic spectral broadening and few-cycle DKS generation, and further into the attosecond regime via dispersive propagation and temporal focusing of the electron \cite{priebeAttosecondElectronPulse2017, ryabovAttosecondMetrologyContinuousbeam2020b, blackNetAccelerationDirect2019a, schonenbergerGenerationCharacterizationAttosecond2019a, yaluninTailoredHighcontrastAttosecond2021}. 
Our work thus opens up new frontiers for ultrafast electron microscopy and ultrafast light-matter interactions driven by chip-based temporal solitons.

\bigskip

%%%%%%%%%%%%%%%%%%%%%%%%%%%%%%%%%%%%%%%%%%%%%%%%%%%%%%%%%%%%%%%%%%%%%%
%%%%%%%%%%%%%%%%%%%%%%%%%%%%%%%%%%%%%%%%%%%%%%%%%%%%%%%%%%%%%%%%%%%%%%
%%%%%%%%%%%%%%%%%%%%%%%%%%%%%%%%%%%%%%%%%%%%%%%%%%%%%%%%%%%%%%%%%%%%%%
\noindent\textbf{Methods}

\smallskip

\begin{footnotesize}

\noindent \textbf{Device design, fabrication and packaging}:
The Si$_3$N$_4$ samples used in this study are air-cladded (top cladding) to allow an interaction of the near field of the intracavity optical states with free electrons. 
The Si$_3$N$_4$ microresonators are fabricated using the photonic Damascene process \cite{pfeifferUltrasmoothSiliconNitride2018b} with an intrinsic linewidth ($\kappa_{0}/2\pi$) of $\sim$ 40 MHz for the quasi-TE mode family. 
The height of the Si$_3$N$_4$ microresonator is optimized to have anomalous dispersion which is an essential requirement to generate bright dissipative Kerr solitons \cite{kippenbergDissipativeKerrSolitons2018} along with significant evanescent electric fields to facilitate electron-light interaction \cite{henkeIntegratedPhotonicsEnables2021}. 
The designed microresonator has dimensions of 2.2 $\mathrm{\mu}$m $\cross$ 800 nm (width $\cross$ height). 
The light is coupled to the microresonator via a bus waveguide of same dimensions to achieve a higher coupling ideality.
In contrast to previous experiments \cite{henkeIntegratedPhotonicsEnables2021, feistCavitymediatedElectronphotonPairs2022a}, the microresonators are designed to have anomalous group velocity dispersion to support the formation of dissipative Kerr soliton (DKS) generation. 
The dispersion of the integrated microresonator can be expressed as a Taylor expansion of cavity modes (denoted by $\mu$) around the pump mode ($\omega_0$): $D_\text{int}(\mu)=\omega_{\mu}-\omega_0-D_1\mu=D_2\mu^2/2+D_3\mu^3/6+...$, where D$_1$ is the first order dispersion related to the free spectral range FSR = $D_1$/2$\pi$ and D$_2$ is the second order dispersion ($D_2$>0 for anomalous dispersion, $D_2$<0 for normal dispersion).
The integrated dispersion and the linewidth of the microresonators are measured using broadband external cavity diode lasers which are calibrated using a fully referenced fiber-based optical frequency comb \cite{delhayeFrequencyCombAssisted2009a}.
The data presented here is mostly acquired using Chip A with wafer ID: D86\_01\_F1\_C16, with FSR = $\SI{100.48}{\GHz}$ and $D_2$ = $\SI{1.21}{\MHz}$. 
We have also performed experiments with several other chips and obtained similar results. 
These chips include: D66\_12\_F16\_C20 with FSR = $\SI{200.54}{\GHz}$ and $D_2$ = $\SI{5.96}{\MHz}$, D66\_12\_F7\_C19 (Chip B, Fig. \ref{Fig:5}c) with FSR = $\SI{200.28}{\GHz}$ and $D_2$ = $\SI{5.50}{\MHz}$, D66\_12\_F18\_C20 with FSR = $\SI{200.56}{\GHz}$ and $D_2$ = $\SI{6.02}{\MHz}$, and D66\_12\_F6\_C20 (Chip C, Fig. \ref{Fig:5}d) with FSR = $\SI{200.27}{\GHz}$ and $D_2$ = $\SI{4.84}{\MHz}$.
The Si$_3$N$_4$ devices are packaged using ultra-high numerical aperture (UHNA) fibers enabling chip-through coupling (fiber-chip-fiber) of around 25\%, and mounted on a custom TEM sample holder. 
The chip is etched into a conical shape to avoid clipping the e-beam and to reduce electrostatic charging of the chip under electron irradiation. 
In addition, a metal layer is deposited on the photonic chip, apart from the Si$_3$N$_4$ waveguide, and vias were etched down to the doped silicon substrate to minimize charging.

\noindent \textbf{Experimental setup}:
The measurements presented in this work were performed at the Göttingen UTEM \cite{feistUltrafastTransmissionElectron2017}, a transmission electron microscope (TEM, JEOL JEM 2100F) based on a Schottky field-emission electron source.
Here, the source is operated in the extended Schottky regime, providing a continuous electron beam of 0.6 eV initial spectral width at a center electron energy of 200 keV. 
In order to reduce the clipping of the electron beam at the extended chip structures the low-magnification scanning TEM mode is used, enabling low electron beam convergence angles and spot sizes < 25 nm depending on the condenser aperture used.
The electron beam is positioned or scanned in front of the resonator structure, mounted on a custom sample holder.
The electron energy spectrum is analyzed using a post-column imaging filter (CEFID, CEOS) as well as a hybrid pixel electron detector based on the Timepix 3 ASIC (EM CheeTah T3, Amsterdam Scientific Instruments). 

We use the transverse electric (TE) mode family throughout this experiment, as it has a high $Q$ and hence a low threshold power ($\propto1/Q^2$), allowing to access the single soliton state. 
For TE polarization, the IELS strength is minimal when the e-beam is positioned tangential to the ring-shaped microresonator and directly above the waveguide, due to a diminished optical field strength along the e-beam direction.
Therefore, we move the e-beam towards the ring center, and the electron path intersects the waveguide near field twice. 
As the electron does not co-propagate with the optical wave for an extended distance in this configuration, electron-photon phase matching \cite{kozakAccelerationSubrelativisticElectrons2017, dahanResonantPhasematchingLight2020, kfirControllingFreeElectrons2020c, henkeIntegratedPhotonicsEnables2021}, namely the matching of the electron group velocity with the optical phase velocity, is less relevant, and we choose a \SI{200}{\keV} electron energy. 

Excitation and characterization of the optical states in the resonator is performed using the setup shown in Fig. \ref{Fig:2}a.
At the input side, it consists of a diode lasers (Toptica CTL 1550), whose frequency is scanned via piezo tuning or single-sideband (SSB) modulation (frequency scanning up to 10 GHz, 150-250 kHz scanning speed) by a sawtooth signal from an arbitrary function generator (AFG).
This laser signal is amplified by an erbium-doped fiber amplifier (EDFA, Keopsys, CEFA-C) and filtered with a tunable bandpass filter of $\sim \SI{0.8}{\nm}$ bandwidth to suppress amplified spontaneous emission. A fiber polarization controller (FPC) is then used to adjust the optical polarization of the amplified CW pump coupled to the TE mode of the \ce{Si3N4} microresonator inside the TEM.
At the output side, the transmitted light is out-coupled from the TEM and analyzed using an optical spectrum analyzer (OSA, Yokogawa AQ6370D) as well as two photodetectors (PD, Newport 1611, 850 MHz bandwidth).
For the latter, optical pump and nonlinearly generated light are separated by a fiber Bragg grating (FBG) and an optical circulator and analyzed individually.
The signals from the photodetectors are monitored on an oscilloscope (OSC, Agilent DSO-X 3034A), while an electronic spectrum analyzer (ESA, Rohde \& Schwarz FPL1007) measures the low-frequency noise of the generated light up to $\sim \SI{1}{\GHz}$  to assess the coherence nature of the nonlinear state (Supplementary Note 10; the microresonator FSR of $\sim \SI{100}{\GHz}$ or $\sim \SI{200}{\GHz}$ is beyond the detection bandwidth).
In addition, an electro-optic modulator (EOM, Thorlabs LN65S) and a vector network analyzer (VNA, Rohde \& Schwarz ZNB 4) are employed to perform modulation probing \cite{guoUniversalDynamicsDeterministic2017} of the intracavity states (cf. Fig. \ref{Fig:4}a).
For transient observation of electron spectral modulation by nonlinear optical states (cf. Fig. \ref{Fig:2}), synchronization is achieved by splitting the AFG signal for both CW pump frequency scanning and electron spectrum acquisition.

\noindent \textbf{Generation of DKS in a TEM}:
To deterministically generate an intracavity optical state, we switch from the continuous scan to a single scan of the pump laser detuning, which is stopped at different points in the stability chart (cf. Fig. \ref{Fig:2}c).
Single-sideband modulation is used to realize a fast frequency scan, which overcomes thermal effects and facilitates DKS generation, as solitons exist on the effectively red-detuned side of the resonance \cite{herrTemporalSolitonsOptical2014c}.
The ability to generate and maintain a DKS state under electron irradiation without any locking techniques during data acquisition (5-10 min) testifies to the radiation hardness of the high-$Q$ \ce{Si3N4} microresonator for applications in harsh environment \cite{braschRadiationHardnessHighQ2014}.

\noindent \textbf{Data acquisition and analysis}:
The fiber packaged, chip-based microresonator is installed in a custom-built holder \cite{henkeIntegratedPhotonicsEnables2021} that allows for a transfer of the optical fibers to the outside and a placement in the sample plane of the TEM.
In a low-magnification scanning TEM mode ($\times$5000 nominal magnification), a focused electron beam of < 25\,nm spot size and low convergence angle ($\SI{40}{\um}$ condenser aperture) is scanned parallel to the chip surface.
The focus of the electron beam is aligned on a small triangle at center height of the resonator ring on the apex of the photonic chip.
Meanwhile, the sample tilt is aligned by maximizing the contrast of the Ramsey fringe pattern in the electron energy spectra for a continuous, single mode optical field in the cavity.
The required characterization of the electron energy distribution is performed with a post-column energy filter, operated in a 64\,eV dispersion mode (energy range $\pm$32\,eV around zero-loss peak), and a hybrid pixel electron detector, resulting in a measurable width of the electron zero-loss peak of $\sim$ 1.1\,eV.
Length scales and beam positions were calibrated via a reference measurement with a sample of known dimensions. 

Scanning the laser frequency via the SSB at an optical input power of about 274\,mW with a 250\,kHz sawtooth signal from an AFG from the blue-detuned side across a quasi-TE mode resonance of the microresonator at approximately 1551.5\,nm, traces of the parametrically generated light are recorded as a function of sweep time and, thus, laser detuning.
Exemplary traces of the blue to red-detuned half cycle of the sweep, presented in Fig. \ref{Fig:2}b, indicate the generation of various nonlinear optical states in the cavity, including the formation of dissipative Kerr solitons.
The electron beam is kept at a fixed position approximately 230\,nm from the resonator surface and 57.3\,$\mu$m from the chip apex throughout this measurement.
In order to reach a sufficient signal-to-noise ratio and high enough electron counts, electron spectra are accumulated over repetitive scans of the pump laser frequency before binning in 8\,ns intervals (acquisition time 10\,s, corresponding to $2.5 \times 10^6$ scans).
The required synchronization of the electron detector to the frequency sweep is achieved by feeding a trigger signal of the AFG to a time-to-digital converter (TDC) of the detector.
The resulting electron energy spectra over the sweep time (Fig. \ref{Fig:2}b) are therefore averaged over different optical states, especially different (multi-)soliton states.
Compensating for an electron beam jitter related to 50\,Hz noise in the laboratory, the center of mass of the electron spectra in each time bin is shifted to coincide with the zero electron energy change. 

For Ramsey interference patterns, the pump laser sweeps (250\,kHz sweep, $\sim$ 280-310\,mW input power) are stopped at different detunings to land in the states described in the stability chart of Fig. \ref{Fig:2}c.
The optical characteristics of the intracavity states are determined from OSA traces (240\,nm span, 0.02\,nm resolution bandwidth) as well as low-frequency RF spectra of the generated light recorded on ESA.
For each optical state, the electron beam is scanned in 300 steps along a 13.8\,$\mu$m line almost parallel to the chip surface (approximately 670\,nm distance) and electron spectra are recorded for each pixel (200\,ms acquisition time).
A section of 30-pixel (1.375\,$\mu$m) length of these line scans are shown in Fig. \ref{Fig:3} along with spectrum line outs at two fixed positions of the Ramsey-type interference pattern.

Changing the pump laser frequency in the single soliton state while monitoring the detuning in the VNA trace enables investigating the impact of single soliton properties, especially the pulse duration and spectral width, on the electron spectra.
Different laser detunings are either set manually in the range of 325-850\,MHz or achieved in a continuous sweep similar to the resonance scans (Supplementary Note 7).
In the former case, e-beam line scans are recorded under the same conditions discussed above while concurrently recording VNA (1.4\,GHz span, 10\,kHz resolution bandwidth) and OSA (240\,nm span, 0.02\,nm resolution bandwidth) traces (cf. Fig. \ref{Fig:4}b\&c), from which the detuning, by comparing the two response maxima, and the soliton pulse duration (cf. Fig. \ref{Fig:4}e), by fitting with a $\sech^2$ function, can be determined. 
Line outs of selected line scans showing electron energy spectra at a fixed position (670\,nm from resonator surface, 62.9\,$\mu$m from chip corner) are plotted in Fig. \ref{Fig:4}d, corresponding to a detuning of 396\,MHz (yellow), 600\,MHz (red) and 857\,MHz (blue).
The maximum coupling strength $|g_\text{max}|$ in Fig. \ref{Fig:4}d, related to the solitons peak electric field strength, is obtained by fitting the electron energy spectra to a model describing the IELS interaction as an average over a distribution of coupling strengths $|g|$ (cf. Eq. \ref{eq:g_def}).
Since the coupling strength is proportional to electric field strength and the continuous electron beam averages over the entire propagating intracavity field, the coupling strength distribution is modeled as hyperbolic secant pulse with amplitude $|g_\text{max}|$ on a constant background of the CW field. 
The electron spectrum is determined by the weighted average of IELS spectra corresponding to the different $|g|$ values in the distribution, which are given by a comb of sidebands of intensity $P_N=J_N(2|g|)^2$ and Gaussian shape \cite{feistQuantumCoherentOptical2015b}.
For the data presented in Fig. \ref{Fig:4}d, the fitting is performed on electron spectra acquired at 100 beam positions spanning over a \SI{4.58}{\um} radial distance, with the dots indicating the average values and the error bars indicating the maximal and minimal values.

The interaction with multi-soliton states is studied by accessing two and three soliton states at a similar detuning (500$\pm$20\,MHz) and recording line scans of the electron energy distribution under the same conditions as above.
Electron energy spectra for different states at a fixed position (670\,nm from resonator surface, 62.9\,$\mu$m from chip corner) are shown in Fig. \ref{Fig:5} alongside OSA traces, from which the number of solitons and their relative position can be determined by fitting the optical spectrum \cite{braschPhotonicChipBased2016c}.
The number of electrons in the spectral plateaus is calculated by integrating the counts of electrons with an energy change higher than \SI{3}{\eV}, at radial positions corresponding to the nodes of the spectral oscillation of the central CW part. 

\noindent \textbf{Theory and simulation}: 
The electron-light interaction is described by the relativistic Dirac equation, which can be reduced to the time-dependent Schr\"{o}dinger equation (TDSE) for a single electron under the nonrecoil approximation, namely, the light-induced electron energy change ($\sim$\SI{100}{\eV}) is much smaller than the electron kinetic energy ($\sim$\SI{100}{\keV}) \cite{garciadeabajoOpticalModulationElectron2021, parkPhotoninducedNearfieldElectron2010}. 
In the velocity gauge,
\begin{equation}
     [\frac{(\hat{\boldsymbol{p}} + e\boldsymbol{A})^2}{2m} - eV]\psi = i \hbar \frac{\partial}{\partial t}\psi.
\end{equation} 
Here, $\hat{\boldsymbol{p}}$ is the momentum operator, $e$ the electron charge, $\boldsymbol{A}$ the vector potential, $m$ the electron mass, $V$ the scalar potential, and $\hbar$ the reduced Planck constant. The electron wavefunction $\psi(\boldsymbol{r},t) = \phi(\boldsymbol{r},t) e^{ik_ez-i\omega_et}$ is a plane wave propagating in the e-beam direction ($\hat{\mathbf{z}}$) with a slowly-varying envelope $\phi(\boldsymbol{r},t)$, an electron momentum $\hbar k_e$, and an electron energy $\hbar \omega_e$. 
With the assumption that the temporal extent of the electron wavefunction is small compared to the characteristic time of the optical field envelope variation, the one-dimensional TDSE can be reduced and then analytically solved by explicit integration, yielding:
\begin{equation}
     \psi (\boldsymbol{r},t) = \phi_0(\boldsymbol{r}) \sum_{N=-\infty}^{+\infty} c_N(x,y)  e^{i(k_e+N\omega_0/v)z - i(\omega_e+N\omega_0)t}.
\end{equation}
Here, $\phi_0(\boldsymbol{r})$ is the initial wavefunction envelope, $\omega_0$ the optical angular frequency, $v$ the electron velocity, $c_N(x,y) = J_N(2|g(x,y)|) (\frac{g(x,y)}{|g(x,y)|})^N$ the amplitude coefficient of the $N$-th photon sideband corresponding to the absorption or emission of integer photon energy $N\hbar\omega_0$, and $J_N$ the Bessel functions of the first kind \cite{dahanResonantPhasematchingLight2020, henkeIntegratedPhotonicsEnables2021}. 
The electron-light coupling parameter,
\begin{equation}\label{eq:g_def}
     g(x,y) = - \frac{e}{2 \hbar \omega_0} \int_{-\infty}^{+\infty} E_z(x,y,z') e^{- i\frac{\omega_0}{v}z'} \,dz',
\end{equation}
is calculated by integrating the product of the electric field $E_z$ in the e-beam direction and the phase matching factor along the e-beam trajectory. 
This theoretical formalism is most accurate for electron interacting with monochromatic CW light, but is also commonly used and validated in the literature for electron interaction with ultrafast light pulses with a finite bandwidth. 
In our experiment, the bandwidth of the frequency comb is relatively narrow compared to the energy uncertainty of the electron, and the field envelope is usually varying slowly in the temporal extent of the electron wavefunction. 
Further more, the interaction region is short and the electron phase modulation can be approximated by the interaction with a monochromatic light field at the place of interaction. 
Hence, we adopt this formalism and obtain a good match between experimental data and simulation results, as shown in the main text. 
A more rigorous theoretical treatment (Supplementary Note 1) obtains similar results as the aforementioned approach.

The intracavity optical field is obtained by solving the Lugiato-Lefever equation (LLE), which amounts to the nonlinear Schr\"{o}dinger equation (NLSE) of optics for a damped, driven, nonlinear Kerr microresonator \cite{kippenbergDissipativeKerrSolitons2018}:
\begin{equation}
     \frac{\partial A}{\partial \tau} - i\frac{D_2}{2}\frac{\partial^2A}{\partial \phi^2} - ig_{\mathrm{0}}|A|^2A + (\kappa/2 + i\Delta)A = \sqrt{\eta \kappa }s_{\mathrm{in}},
\end{equation}
where $\phi$ is the angular coordinate of the microresonator in a rotating frame at the group velocity, $\tau$ is the slow time, $A(\phi,\tau)$ is the intracavity optical field envelope (with the instantaneous photon number determined by $\langle \hat n_\mathrm{t}\rangle = |A|^2$), $D_2$ is the group velocity dispersion (GVD), and $\Delta$ is the pump detuning. 
The input field flux $s_{\mathrm{in}}$ is associated with the optical pump power by $P = \hbar \omega_0 |s_{\mathrm{in}}|^2$. 
The coupling efficiency $\eta = \kappa_\mathrm{ex} / \kappa$, with the external coupling rate $\kappa_\mathrm{ex}$ and the total loss rate $\kappa$. 
The single-photon Kerr frequency shift is $g_{\mathrm{0}} = \hbar \omega_0^2 c n_2 / n^2 V_0$, with the speed of light $c$, the linear refractive index $n$, the nonlinear refractive index $n_2$, and the nonlinear effective mode volume $V_0$. 
The transverse optical mode profile $\mathbf{e}(x,y)$ (in particular the tangential field $e_\phi$ and radial field $e_r$) is obtained by solving Maxwell's equations with linear constitutive relations. 
Taking into consideration the intracavity field envelope and the optical polarization, the electron-light coupling parameter is
\begin{equation} \label{eq:g_cav}
    \begin{aligned}
             g(x,y) = - \frac{e}{2 \hbar \omega_0} &\int_{-\infty}^{+\infty} A(\phi+\phi_0+D_1\frac{z'}{v},\tau)(-e_\phi(x,y)\cos\phi\\
             &-e_r(x,y)\sin\phi)e^{-in\frac{\omega_0}{c}R\phi}e^{- i\frac{\omega_0}{v}z'} \,dz',
    \end{aligned}
\end{equation}
where $D_1/2\pi$ is the microresonator free spectral range (FSR), $R$ is the microresonator ring radius, and $\phi_0$ accounts for the angular offset of the rotating frame from the laboratory frame.

For numerical simulations, the intracavity field envelope is obtained by numerically solving the LLE with a split-step method, with 512 frequency modes considered. 
For stable intracavity states (e.g. CW, Turing pattern, stable DKS), the detuning-dependent field envelope is obtained by scanning the pump frequency from far blue-detuned (\SI{-0.5}{\GHz}) to far red-detuned (\SI{2.5}{\GHz}), and specific intracavity waveforms are selected by referring to the experimentally measured optical spectra and pump detuning. 
Due to the stochastic nature of DKS generation, the simulation is repeated 300 times to achieve the desired DKS states (in particular the single soliton state) that match with experimental data. 
For chaotic states (e.g. chaotic MI), the solver is initially hard seeded with the solution of the corresponding states taken from the previous detuning-dependent simulation results, and then the time-dependent intracavity waveforms are simulated by propagating the slow time for \SI{10}{\us} with a step size of \SI{100}{\ps}. 
The optical near field mode profile is obtained from a finite-element-method electromagnetic solver (COMSOL Multiphysics) with a mode analysis of the axial symmetric waveguide cross-section. 
The electron-light coupling parameter is then calculated via Eq. (\ref{eq:g_cav}).
For stable states, the dependence on the slow time can be dropped, and the intensity of the $N$-th photon sideband of the electron spectrum is calculated by averaging over the angular offset: $I_{N,\mathrm{stable}}(x,y) = \langle|J_N(2|g(x,y)|)|^2\rangle_{\phi_0}$, due to the random arrival time of electrons in a continuous beam. 
On the other hand, for unstable states, the average is taken over the slow time: $I_{N,\mathrm{unstable}}(x,y) = \langle|J_N(2|g(x,y)|)|^2\rangle_{\tau}$, as the intracavity field is chaotic and vastly varying. 
To get the final simulated electron spectrum, the photon sideband intensities are convoluted with a Gaussian function that represents the experimental ZLP width and spectrometer resolution. 
All simulation parameters are taken from experimental measurement (whenever available) or design specifications, including the electron energy, pump power, pump frequency, pump detuning, FSR, GVD, waveguide and microresonator dimensions, as well as the external coupling and the internal loss rates. 
For a closer match between the simulation and the experiment, the e-beam position in the simulation is slight adjusted (by less than \SI{400}{\nm}) to account for the experimental uncertainty of e-beam position, in-coupled optical power, exact soliton existence range and effective detuning, and the microresonator geometry from fabrication process variations. 
The unique radial position dependence of the Turing pattern Ramsey interference also assists in determining the approximate e-beam position (Supplementary Note 2).

As a final remark, we have benchmarked the above theoretical approach against the more rigorous treatment that considers a non-uniform modulation of the electron wavefunction imposed by light with a varying amplitude (e.g. the DKS pulse), and obtained almost identical results (Supplementary Note 1). 
This corroborates the validity of our theoretical approach without the need for correction from non-uniform modulation, mainly due to time-averaging of the electron spectra and the long DKS pulse duration compared to the temporal extent of the electron wavefunction. 
In future studies involving few-optical-cycle DKS pulses and broadband (e.g. octave spanning) optical frequency combs, the more rigorous approach should be adopted.

\medskip

\noindent \textbf{Acknowledgments}:
This material is based on work supported by the Air Force Office of Scientific Research under award FA9550-19-1-0250 and by the Swiss National Science Foundation under grant agreement 185870 (Ambizione). Y.Y. acknowledges support from the EU H2020 research and innovation program under the Marie Skłodowska-Curie IF grant agreement 101033593 (SEPhIM). All photonic integrated circuit samples were fabricated in the Center of MicroNanoTechnology (CMi) and the Institute of Physics cleanroom at EPFL. The experiments were conducted at the Göttingen UTEM Lab, funded by the Deutsche Forschungsgemeinschaft (DFG, German Research Foundation) through grant number 432680300/SFB 1456 (project C01) and the Gottfried Wilhelm Leibniz programme, and the EU H2020 research and innovation programme under grant agreement number 101017720 (FET-Proactive EBEAM). 

\medskip

\noindent \textbf{Author contribution}:
Conceptualization, Supervision, Project administration, Funding acquisition: T.J.K., C.R.
Methodology: A.S.R., Y.Y., J.-W.H., F.J.K., G.H., Z.Q., A.F., R.N.W., A.Tu., A.Ti.
Software: F.J.K., A.Ti., J.-W.H., A.Tu.
Validation: A.S.R., J.-W.H., Y.Y., F.J.K.
Formal analysis: Y.Y., J.-W.H., A.S.R., F.J.K., G.H.
Investigation: J.-W.H., Y.Y., F.J.K., A.S.R., G.H., G.A., A.F.
Resources: F.J.K., Z.Q., R.N.W., A.F., G.A.
Data curation: J.-W.H., F.J.K., Y.Y., A.S.R.
Visualization: Y.Y., F.J.K., A.S.R., J.-W.H.
Writing (original draft preparation): Y.Y., J.-W.H., A.S.R., F.J.K., C.R., T.J.K.
Writing (review and editing): all authors.

\medskip

\noindent \textbf{Data Availability Statement}: The code and data used to produce the plots within this work will be released on the repository \texttt{Zenodo} upon publication of this preprint.

\end{footnotesize}

\renewcommand{\bibpreamble}{
$^\ast$These authors contributed equally to this work.\\
$^\dag$\textcolor{magenta}{yujia.yang@epfl.ch}\\
$^\ddag$\textcolor{magenta}{claus.ropers@mpinat.mpg.de}\\
$^\S$\textcolor{magenta}{tobias.kippenberg@epfl.ch}\\
}
\pretolerance=0
\bigskip
\bibliographystyle{apsrev4-2}
\bibliography{bibliography}

%apsrev4-2.bst 2019-01-14 (MD) hand-edited version of apsrev4-1.bst
%Control: key (0)
%Control: author (72) initials jnrlst
%Control: editor formatted (1) identically to author
%Control: production of article title (-1) disabled
%Control: page (0) single
%Control: year (1) truncated
%Control: production of eprint (0) enabled
\begin{thebibliography}{63}%
\makeatletter
\providecommand \@ifxundefined [1]{%
 \@ifx{#1\undefined}
}%
\providecommand \@ifnum [1]{%
 \ifnum #1\expandafter \@firstoftwo
 \else \expandafter \@secondoftwo
 \fi
}%
\providecommand \@ifx [1]{%
 \ifx #1\expandafter \@firstoftwo
 \else \expandafter \@secondoftwo
 \fi
}%
\providecommand \natexlab [1]{#1}%
\providecommand \enquote  [1]{``#1''}%
\providecommand \bibnamefont  [1]{#1}%
\providecommand \bibfnamefont [1]{#1}%
\providecommand \citenamefont [1]{#1}%
\providecommand \href@noop [0]{\@secondoftwo}%
\providecommand \href [0]{\begingroup \@sanitize@url \@href}%
\providecommand \@href[1]{\@@startlink{#1}\@@href}%
\providecommand \@@href[1]{\endgroup#1\@@endlink}%
\providecommand \@sanitize@url [0]{\catcode `\\12\catcode `\$12\catcode
  `\&12\catcode `\#12\catcode `\^12\catcode `\_12\catcode `\%12\relax}%
\providecommand \@@startlink[1]{}%
\providecommand \@@endlink[0]{}%
\providecommand \url  [0]{\begingroup\@sanitize@url \@url }%
\providecommand \@url [1]{\endgroup\@href {#1}{\urlprefix }}%
\providecommand \urlprefix  [0]{URL }%
\providecommand \Eprint [0]{\href }%
\providecommand \doibase [0]{https://doi.org/}%
\providecommand \selectlanguage [0]{\@gobble}%
\providecommand \bibinfo  [0]{\@secondoftwo}%
\providecommand \bibfield  [0]{\@secondoftwo}%
\providecommand \translation [1]{[#1]}%
\providecommand \BibitemOpen [0]{}%
\providecommand \bibitemStop [0]{}%
\providecommand \bibitemNoStop [0]{.\EOS\space}%
\providecommand \EOS [0]{\spacefactor3000\relax}%
\providecommand \BibitemShut  [1]{\csname bibitem#1\endcsname}%
\let\auto@bib@innerbib\@empty
%</preamble>
\bibitem [{\citenamefont
  {Ruska}(1987)}]{ruskaDevelopmentElectronMicroscope1987}%
  \BibitemOpen
  \bibfield  {author} {\bibinfo {author} {\bibfnamefont {E.}~\bibnamefont
  {Ruska}},\ }\href {https://doi.org/10.1103/RevModPhys.59.627} {\bibfield
  {journal} {\bibinfo  {journal} {Reviews of Modern Physics}\ }\textbf
  {\bibinfo {volume} {59}},\ \bibinfo {pages} {627} (\bibinfo {year}
  {1987})}\BibitemShut {NoStop}%
\bibitem [{\citenamefont
  {Dubochet}(2018)}]{dubochetDevelopmentElectronCryoMicroscopy2018}%
  \BibitemOpen
  \bibfield  {author} {\bibinfo {author} {\bibfnamefont {J.}~\bibnamefont
  {Dubochet}},\ }\href {https://doi.org/10.1002/anie.201804280} {\bibfield
  {journal} {\bibinfo  {journal} {Angewandte Chemie International Edition}\
  }\textbf {\bibinfo {volume} {57}},\ \bibinfo {pages} {10842} (\bibinfo {year}
  {2018})}\BibitemShut {NoStop}%
\bibitem [{\citenamefont {Barwick}\ \emph {et~al.}(2009)\citenamefont
  {Barwick}, \citenamefont {Flannigan},\ and\ \citenamefont
  {Zewail}}]{barwickPhotoninducedNearfieldElectron2009a}%
  \BibitemOpen
  \bibfield  {author} {\bibinfo {author} {\bibfnamefont {B.}~\bibnamefont
  {Barwick}}, \bibinfo {author} {\bibfnamefont {D.~J.}\ \bibnamefont
  {Flannigan}},\ and\ \bibinfo {author} {\bibfnamefont {A.~H.}\ \bibnamefont
  {Zewail}},\ }\href {https://doi.org/10.1038/nature08662} {\bibfield
  {journal} {\bibinfo  {journal} {Nature}\ }\textbf {\bibinfo {volume} {462}},\
  \bibinfo {pages} {902} (\bibinfo {year} {2009})}\BibitemShut {NoStop}%
\bibitem [{\citenamefont {Ryabov}\ and\ \citenamefont
  {Baum}(2016)}]{ryabovElectronMicroscopyElectromagnetic2016b}%
  \BibitemOpen
  \bibfield  {author} {\bibinfo {author} {\bibfnamefont {A.}~\bibnamefont
  {Ryabov}}\ and\ \bibinfo {author} {\bibfnamefont {P.}~\bibnamefont {Baum}},\
  }\href {https://doi.org/10.1126/science.aaf8589} {\bibfield  {journal}
  {\bibinfo  {journal} {Science}\ }\textbf {\bibinfo {volume} {353}},\ \bibinfo
  {pages} {374} (\bibinfo {year} {2016})}\BibitemShut {NoStop}%
\bibitem [{\citenamefont {Polman}\ \emph {et~al.}(2019)\citenamefont {Polman},
  \citenamefont {Kociak},\ and\ \citenamefont {{Garc{\'i}a de
  Abajo}}}]{polmanElectronbeamSpectroscopyNanophotonics2019a}%
  \BibitemOpen
  \bibfield  {author} {\bibinfo {author} {\bibfnamefont {A.}~\bibnamefont
  {Polman}}, \bibinfo {author} {\bibfnamefont {M.}~\bibnamefont {Kociak}},\
  and\ \bibinfo {author} {\bibfnamefont {F.~J.}\ \bibnamefont {{Garc{\'i}a de
  Abajo}}},\ }\href {https://doi.org/10.1038/s41563-019-0409-1} {\bibfield
  {journal} {\bibinfo  {journal} {Nature Materials}\ }\textbf {\bibinfo
  {volume} {18}},\ \bibinfo {pages} {1158} (\bibinfo {year}
  {2019})}\BibitemShut {NoStop}%
\bibitem [{\citenamefont {Wang}\ \emph {et~al.}(2020)\citenamefont {Wang},
  \citenamefont {Dahan}, \citenamefont {Shentcis}, \citenamefont {Kauffmann},
  \citenamefont {Ben~Hayun}, \citenamefont {Reinhardt}, \citenamefont
  {Tsesses},\ and\ \citenamefont {Kaminer}}]{wangCoherentInteractionFree2020e}%
  \BibitemOpen
  \bibfield  {author} {\bibinfo {author} {\bibfnamefont {K.}~\bibnamefont
  {Wang}}, \bibinfo {author} {\bibfnamefont {R.}~\bibnamefont {Dahan}},
  \bibinfo {author} {\bibfnamefont {M.}~\bibnamefont {Shentcis}}, \bibinfo
  {author} {\bibfnamefont {Y.}~\bibnamefont {Kauffmann}}, \bibinfo {author}
  {\bibfnamefont {A.}~\bibnamefont {Ben~Hayun}}, \bibinfo {author}
  {\bibfnamefont {O.}~\bibnamefont {Reinhardt}}, \bibinfo {author}
  {\bibfnamefont {S.}~\bibnamefont {Tsesses}},\ and\ \bibinfo {author}
  {\bibfnamefont {I.}~\bibnamefont {Kaminer}},\ }\href
  {https://doi.org/10.1038/s41586-020-2321-x} {\bibfield  {journal} {\bibinfo
  {journal} {Nature}\ }\textbf {\bibinfo {volume} {582}},\ \bibinfo {pages}
  {50} (\bibinfo {year} {2020})}\BibitemShut {NoStop}%
\bibitem [{\citenamefont {Kfir}\ \emph {et~al.}(2020)\citenamefont {Kfir},
  \citenamefont {{Louren{\c c}o-Martins}}, \citenamefont {Storeck},
  \citenamefont {Sivis}, \citenamefont {Harvey}, \citenamefont {Kippenberg},
  \citenamefont {Feist},\ and\ \citenamefont
  {Ropers}}]{kfirControllingFreeElectrons2020c}%
  \BibitemOpen
  \bibfield  {author} {\bibinfo {author} {\bibfnamefont {O.}~\bibnamefont
  {Kfir}}, \bibinfo {author} {\bibfnamefont {H.}~\bibnamefont {{Louren{\c
  c}o-Martins}}}, \bibinfo {author} {\bibfnamefont {G.}~\bibnamefont
  {Storeck}}, \bibinfo {author} {\bibfnamefont {M.}~\bibnamefont {Sivis}},
  \bibinfo {author} {\bibfnamefont {T.~R.}\ \bibnamefont {Harvey}}, \bibinfo
  {author} {\bibfnamefont {T.~J.}\ \bibnamefont {Kippenberg}}, \bibinfo
  {author} {\bibfnamefont {A.}~\bibnamefont {Feist}},\ and\ \bibinfo {author}
  {\bibfnamefont {C.}~\bibnamefont {Ropers}},\ }\href
  {https://doi.org/10.1038/s41586-020-2320-y} {\bibfield  {journal} {\bibinfo
  {journal} {Nature}\ }\textbf {\bibinfo {volume} {582}},\ \bibinfo {pages}
  {46} (\bibinfo {year} {2020})}\BibitemShut {NoStop}%
\bibitem [{\citenamefont {Feist}\ \emph {et~al.}(2015)\citenamefont {Feist},
  \citenamefont {Echternkamp}, \citenamefont {Schauss}, \citenamefont
  {Yalunin}, \citenamefont {Sch{\"a}fer},\ and\ \citenamefont
  {Ropers}}]{feistQuantumCoherentOptical2015b}%
  \BibitemOpen
  \bibfield  {author} {\bibinfo {author} {\bibfnamefont {A.}~\bibnamefont
  {Feist}}, \bibinfo {author} {\bibfnamefont {K.~E.}\ \bibnamefont
  {Echternkamp}}, \bibinfo {author} {\bibfnamefont {J.}~\bibnamefont
  {Schauss}}, \bibinfo {author} {\bibfnamefont {S.~V.}\ \bibnamefont
  {Yalunin}}, \bibinfo {author} {\bibfnamefont {S.}~\bibnamefont
  {Sch{\"a}fer}},\ and\ \bibinfo {author} {\bibfnamefont {C.}~\bibnamefont
  {Ropers}},\ }\href {https://doi.org/10.1038/nature14463} {\bibfield
  {journal} {\bibinfo  {journal} {Nature}\ }\textbf {\bibinfo {volume} {521}},\
  \bibinfo {pages} {200} (\bibinfo {year} {2015})}\BibitemShut {NoStop}%
\bibitem [{\citenamefont {Henke}\ \emph {et~al.}(2021)\citenamefont {Henke},
  \citenamefont {Raja}, \citenamefont {Feist}, \citenamefont {Huang},
  \citenamefont {Arend}, \citenamefont {Yang}, \citenamefont {Kappert},
  \citenamefont {Wang}, \citenamefont {M{\"o}ller}, \citenamefont {Pan},
  \citenamefont {Liu}, \citenamefont {Kfir}, \citenamefont {Ropers},\ and\
  \citenamefont {Kippenberg}}]{henkeIntegratedPhotonicsEnables2021}%
  \BibitemOpen
  \bibfield  {author} {\bibinfo {author} {\bibfnamefont {J.-W.}\ \bibnamefont
  {Henke}}, \bibinfo {author} {\bibfnamefont {A.~S.}\ \bibnamefont {Raja}},
  \bibinfo {author} {\bibfnamefont {A.}~\bibnamefont {Feist}}, \bibinfo
  {author} {\bibfnamefont {G.}~\bibnamefont {Huang}}, \bibinfo {author}
  {\bibfnamefont {G.}~\bibnamefont {Arend}}, \bibinfo {author} {\bibfnamefont
  {Y.}~\bibnamefont {Yang}}, \bibinfo {author} {\bibfnamefont {F.~J.}\
  \bibnamefont {Kappert}}, \bibinfo {author} {\bibfnamefont {R.~N.}\
  \bibnamefont {Wang}}, \bibinfo {author} {\bibfnamefont {M.}~\bibnamefont
  {M{\"o}ller}}, \bibinfo {author} {\bibfnamefont {J.}~\bibnamefont {Pan}},
  \bibinfo {author} {\bibfnamefont {J.}~\bibnamefont {Liu}}, \bibinfo {author}
  {\bibfnamefont {O.}~\bibnamefont {Kfir}}, \bibinfo {author} {\bibfnamefont
  {C.}~\bibnamefont {Ropers}},\ and\ \bibinfo {author} {\bibfnamefont {T.~J.}\
  \bibnamefont {Kippenberg}},\ }\href
  {https://doi.org/10.1038/s41586-021-04197-5} {\bibfield  {journal} {\bibinfo
  {journal} {Nature}\ }\textbf {\bibinfo {volume} {600}},\ \bibinfo {pages}
  {653} (\bibinfo {year} {2021})}\BibitemShut {NoStop}%
\bibitem [{\citenamefont {Priebe}\ \emph {et~al.}(2017)\citenamefont {Priebe},
  \citenamefont {Rathje}, \citenamefont {Yalunin}, \citenamefont {Hohage},
  \citenamefont {Feist}, \citenamefont {Sch{\"a}fer},\ and\ \citenamefont
  {Ropers}}]{priebeAttosecondElectronPulse2017}%
  \BibitemOpen
  \bibfield  {author} {\bibinfo {author} {\bibfnamefont {K.~E.}\ \bibnamefont
  {Priebe}}, \bibinfo {author} {\bibfnamefont {C.}~\bibnamefont {Rathje}},
  \bibinfo {author} {\bibfnamefont {S.~V.}\ \bibnamefont {Yalunin}}, \bibinfo
  {author} {\bibfnamefont {T.}~\bibnamefont {Hohage}}, \bibinfo {author}
  {\bibfnamefont {A.}~\bibnamefont {Feist}}, \bibinfo {author} {\bibfnamefont
  {S.}~\bibnamefont {Sch{\"a}fer}},\ and\ \bibinfo {author} {\bibfnamefont
  {C.}~\bibnamefont {Ropers}},\ }\href
  {https://doi.org/10.1038/s41566-017-0045-8} {\bibfield  {journal} {\bibinfo
  {journal} {Nature Photonics}\ }\textbf {\bibinfo {volume} {11}},\ \bibinfo
  {pages} {793} (\bibinfo {year} {2017})}\BibitemShut {NoStop}%
\bibitem [{\citenamefont {Morimoto}\ and\ \citenamefont
  {Baum}(2018)}]{morimotoDiffractionMicroscopyAttosecond2018}%
  \BibitemOpen
  \bibfield  {author} {\bibinfo {author} {\bibfnamefont {Y.}~\bibnamefont
  {Morimoto}}\ and\ \bibinfo {author} {\bibfnamefont {P.}~\bibnamefont
  {Baum}},\ }\href {https://doi.org/10.1038/s41567-017-0007-6} {\bibfield
  {journal} {\bibinfo  {journal} {Nature Physics}\ }\textbf {\bibinfo {volume}
  {14}},\ \bibinfo {pages} {252} (\bibinfo {year} {2018})}\BibitemShut
  {NoStop}%
\bibitem [{\citenamefont {Ryabov}\ \emph {et~al.}(2020)\citenamefont {Ryabov},
  \citenamefont {Thurner}, \citenamefont {Nabben}, \citenamefont {Tsarev},\
  and\ \citenamefont {Baum}}]{ryabovAttosecondMetrologyContinuousbeam2020b}%
  \BibitemOpen
  \bibfield  {author} {\bibinfo {author} {\bibfnamefont {A.}~\bibnamefont
  {Ryabov}}, \bibinfo {author} {\bibfnamefont {J.~W.}\ \bibnamefont {Thurner}},
  \bibinfo {author} {\bibfnamefont {D.}~\bibnamefont {Nabben}}, \bibinfo
  {author} {\bibfnamefont {M.~V.}\ \bibnamefont {Tsarev}},\ and\ \bibinfo
  {author} {\bibfnamefont {P.}~\bibnamefont {Baum}},\ }\href
  {https://doi.org/10.1126/sciadv.abb1393} {\bibfield  {journal} {\bibinfo
  {journal} {Science Advances}\ }\textbf {\bibinfo {volume} {6}},\ \bibinfo
  {pages} {eabb1393} (\bibinfo {year} {2020})}\BibitemShut {NoStop}%
\bibitem [{\citenamefont {{Garc{\'i}a de Abajo}}\ \emph
  {et~al.}(2016)\citenamefont {{Garc{\'i}a de Abajo}}, \citenamefont
  {Barwick},\ and\ \citenamefont
  {Carbone}}]{garciadeabajoElectronDiffractionPlasmon2016}%
  \BibitemOpen
  \bibfield  {author} {\bibinfo {author} {\bibfnamefont {F.~J.}\ \bibnamefont
  {{Garc{\'i}a de Abajo}}}, \bibinfo {author} {\bibfnamefont {B.}~\bibnamefont
  {Barwick}},\ and\ \bibinfo {author} {\bibfnamefont {F.}~\bibnamefont
  {Carbone}},\ }\href {https://doi.org/10.1103/PhysRevB.94.041404} {\bibfield
  {journal} {\bibinfo  {journal} {Physical Review B}\ }\textbf {\bibinfo
  {volume} {94}},\ \bibinfo {pages} {041404} (\bibinfo {year}
  {2016})}\BibitemShut {NoStop}%
\bibitem [{\citenamefont {Vanacore}\ \emph {et~al.}(2019)\citenamefont
  {Vanacore}, \citenamefont {Berruto}, \citenamefont {Madan}, \citenamefont
  {Pomarico}, \citenamefont {Biagioni}, \citenamefont {Lamb}, \citenamefont
  {McGrouther}, \citenamefont {Reinhardt}, \citenamefont {Kaminer},
  \citenamefont {Barwick}, \citenamefont {Larocque}, \citenamefont {Grillo},
  \citenamefont {Karimi}, \citenamefont {{Garc{\'i}a de Abajo}},\ and\
  \citenamefont {Carbone}}]{vanacoreUltrafastGenerationControl2019b}%
  \BibitemOpen
  \bibfield  {author} {\bibinfo {author} {\bibfnamefont {G.~M.}\ \bibnamefont
  {Vanacore}}, \bibinfo {author} {\bibfnamefont {G.}~\bibnamefont {Berruto}},
  \bibinfo {author} {\bibfnamefont {I.}~\bibnamefont {Madan}}, \bibinfo
  {author} {\bibfnamefont {E.}~\bibnamefont {Pomarico}}, \bibinfo {author}
  {\bibfnamefont {P.}~\bibnamefont {Biagioni}}, \bibinfo {author}
  {\bibfnamefont {R.~J.}\ \bibnamefont {Lamb}}, \bibinfo {author}
  {\bibfnamefont {D.}~\bibnamefont {McGrouther}}, \bibinfo {author}
  {\bibfnamefont {O.}~\bibnamefont {Reinhardt}}, \bibinfo {author}
  {\bibfnamefont {I.}~\bibnamefont {Kaminer}}, \bibinfo {author} {\bibfnamefont
  {B.}~\bibnamefont {Barwick}}, \bibinfo {author} {\bibfnamefont
  {H.}~\bibnamefont {Larocque}}, \bibinfo {author} {\bibfnamefont
  {V.}~\bibnamefont {Grillo}}, \bibinfo {author} {\bibfnamefont
  {E.}~\bibnamefont {Karimi}}, \bibinfo {author} {\bibfnamefont {F.~J.}\
  \bibnamefont {{Garc{\'i}a de Abajo}}},\ and\ \bibinfo {author} {\bibfnamefont
  {F.}~\bibnamefont {Carbone}},\ }\href
  {https://doi.org/10.1038/s41563-019-0336-1} {\bibfield  {journal} {\bibinfo
  {journal} {Nature Materials}\ }\textbf {\bibinfo {volume} {18}},\ \bibinfo
  {pages} {573} (\bibinfo {year} {2019})}\BibitemShut {NoStop}%
\bibitem [{\citenamefont {Kone{\v c}n{\'a}}\ and\ \citenamefont {{de
  Abajo}}(2020)}]{konecnaElectronBeamAberration2020}%
  \BibitemOpen
  \bibfield  {author} {\bibinfo {author} {\bibfnamefont {A.}~\bibnamefont
  {Kone{\v c}n{\'a}}}\ and\ \bibinfo {author} {\bibfnamefont {F.~J.~G.}\
  \bibnamefont {{de Abajo}}},\ }\href
  {https://doi.org/10.1103/PhysRevLett.125.030801} {\bibfield  {journal}
  {\bibinfo  {journal} {Physical Review Letters}\ }\textbf {\bibinfo {volume}
  {125}},\ \bibinfo {pages} {030801} (\bibinfo {year} {2020})}\BibitemShut
  {NoStop}%
\bibitem [{\citenamefont {Feist}\ \emph {et~al.}(2020)\citenamefont {Feist},
  \citenamefont {Yalunin}, \citenamefont {Sch{\"a}fer},\ and\ \citenamefont
  {Ropers}}]{feistHighpurityFreeelectronMomentum2020a}%
  \BibitemOpen
  \bibfield  {author} {\bibinfo {author} {\bibfnamefont {A.}~\bibnamefont
  {Feist}}, \bibinfo {author} {\bibfnamefont {S.~V.}\ \bibnamefont {Yalunin}},
  \bibinfo {author} {\bibfnamefont {S.}~\bibnamefont {Sch{\"a}fer}},\ and\
  \bibinfo {author} {\bibfnamefont {C.}~\bibnamefont {Ropers}},\ }\href
  {https://doi.org/10.1103/PhysRevResearch.2.043227} {\bibfield  {journal}
  {\bibinfo  {journal} {Physical Review Research}\ }\textbf {\bibinfo {volume}
  {2}},\ \bibinfo {pages} {043227} (\bibinfo {year} {2020})}\BibitemShut
  {NoStop}%
\bibitem [{\citenamefont {Madan}\ \emph {et~al.}(2022)\citenamefont {Madan},
  \citenamefont {Leccese}, \citenamefont {Mazur}, \citenamefont {Barantani},
  \citenamefont {LaGrange}, \citenamefont {Sapozhnik}, \citenamefont {Tengdin},
  \citenamefont {Gargiulo}, \citenamefont {Rotunno}, \citenamefont {Olaya},
  \citenamefont {Kaminer}, \citenamefont {Grillo}, \citenamefont {{de Abajo}},
  \citenamefont {Carbone},\ and\ \citenamefont
  {Vanacore}}]{madanUltrafastTransverseModulation2022}%
  \BibitemOpen
  \bibfield  {author} {\bibinfo {author} {\bibfnamefont {I.}~\bibnamefont
  {Madan}}, \bibinfo {author} {\bibfnamefont {V.}~\bibnamefont {Leccese}},
  \bibinfo {author} {\bibfnamefont {A.}~\bibnamefont {Mazur}}, \bibinfo
  {author} {\bibfnamefont {F.}~\bibnamefont {Barantani}}, \bibinfo {author}
  {\bibfnamefont {T.}~\bibnamefont {LaGrange}}, \bibinfo {author}
  {\bibfnamefont {A.}~\bibnamefont {Sapozhnik}}, \bibinfo {author}
  {\bibfnamefont {P.~M.}\ \bibnamefont {Tengdin}}, \bibinfo {author}
  {\bibfnamefont {S.}~\bibnamefont {Gargiulo}}, \bibinfo {author}
  {\bibfnamefont {E.}~\bibnamefont {Rotunno}}, \bibinfo {author} {\bibfnamefont
  {J.-C.}\ \bibnamefont {Olaya}}, \bibinfo {author} {\bibfnamefont
  {I.}~\bibnamefont {Kaminer}}, \bibinfo {author} {\bibfnamefont
  {V.}~\bibnamefont {Grillo}}, \bibinfo {author} {\bibfnamefont {F.~J.~G.}\
  \bibnamefont {{de Abajo}}}, \bibinfo {author} {\bibfnamefont
  {F.}~\bibnamefont {Carbone}},\ and\ \bibinfo {author} {\bibfnamefont {G.~M.}\
  \bibnamefont {Vanacore}},\ }\href
  {https://doi.org/10.1021/acsphotonics.2c00850} {\bibfield  {journal}
  {\bibinfo  {journal} {ACS Photonics}\ }\textbf {\bibinfo {volume} {9}},\
  \bibinfo {pages} {3215} (\bibinfo {year} {2022})}\BibitemShut {NoStop}%
\bibitem [{\citenamefont {England}\ \emph {et~al.}(2014)\citenamefont
  {England}, \citenamefont {Noble}, \citenamefont {Bane}, \citenamefont
  {Dowell}, \citenamefont {Ng}, \citenamefont {Spencer}, \citenamefont
  {Tantawi}, \citenamefont {Wu}, \citenamefont {Byer}, \citenamefont {Peralta},
  \citenamefont {Soong}, \citenamefont {Chang}, \citenamefont {Montazeri},
  \citenamefont {Wolf}, \citenamefont {Cowan}, \citenamefont {Dawson},
  \citenamefont {Gai}, \citenamefont {Hommelhoff}, \citenamefont {Huang},
  \citenamefont {Jing}, \citenamefont {McGuinness}, \citenamefont {Palmer},
  \citenamefont {Naranjo}, \citenamefont {Rosenzweig}, \citenamefont {Travish},
  \citenamefont {Mizrahi}, \citenamefont {Schachter}, \citenamefont {Sears},
  \citenamefont {Werner},\ and\ \citenamefont
  {Yoder}}]{englandDielectricLaserAccelerators2014a}%
  \BibitemOpen
  \bibfield  {author} {\bibinfo {author} {\bibfnamefont {R.~J.}\ \bibnamefont
  {England}}, \bibinfo {author} {\bibfnamefont {R.~J.}\ \bibnamefont {Noble}},
  \bibinfo {author} {\bibfnamefont {K.}~\bibnamefont {Bane}}, \bibinfo {author}
  {\bibfnamefont {D.~H.}\ \bibnamefont {Dowell}}, \bibinfo {author}
  {\bibfnamefont {C.-K.}\ \bibnamefont {Ng}}, \bibinfo {author} {\bibfnamefont
  {J.~E.}\ \bibnamefont {Spencer}}, \bibinfo {author} {\bibfnamefont
  {S.}~\bibnamefont {Tantawi}}, \bibinfo {author} {\bibfnamefont
  {Z.}~\bibnamefont {Wu}}, \bibinfo {author} {\bibfnamefont {R.~L.}\
  \bibnamefont {Byer}}, \bibinfo {author} {\bibfnamefont {E.}~\bibnamefont
  {Peralta}}, \bibinfo {author} {\bibfnamefont {K.}~\bibnamefont {Soong}},
  \bibinfo {author} {\bibfnamefont {C.-M.}\ \bibnamefont {Chang}}, \bibinfo
  {author} {\bibfnamefont {B.}~\bibnamefont {Montazeri}}, \bibinfo {author}
  {\bibfnamefont {S.~J.}\ \bibnamefont {Wolf}}, \bibinfo {author}
  {\bibfnamefont {B.}~\bibnamefont {Cowan}}, \bibinfo {author} {\bibfnamefont
  {J.}~\bibnamefont {Dawson}}, \bibinfo {author} {\bibfnamefont
  {W.}~\bibnamefont {Gai}}, \bibinfo {author} {\bibfnamefont {P.}~\bibnamefont
  {Hommelhoff}}, \bibinfo {author} {\bibfnamefont {Y.-C.}\ \bibnamefont
  {Huang}}, \bibinfo {author} {\bibfnamefont {C.}~\bibnamefont {Jing}},
  \bibinfo {author} {\bibfnamefont {C.}~\bibnamefont {McGuinness}}, \bibinfo
  {author} {\bibfnamefont {R.~B.}\ \bibnamefont {Palmer}}, \bibinfo {author}
  {\bibfnamefont {B.}~\bibnamefont {Naranjo}}, \bibinfo {author} {\bibfnamefont
  {J.}~\bibnamefont {Rosenzweig}}, \bibinfo {author} {\bibfnamefont
  {G.}~\bibnamefont {Travish}}, \bibinfo {author} {\bibfnamefont
  {A.}~\bibnamefont {Mizrahi}}, \bibinfo {author} {\bibfnamefont
  {L.}~\bibnamefont {Schachter}}, \bibinfo {author} {\bibfnamefont
  {C.}~\bibnamefont {Sears}}, \bibinfo {author} {\bibfnamefont {G.~R.}\
  \bibnamefont {Werner}},\ and\ \bibinfo {author} {\bibfnamefont {R.~B.}\
  \bibnamefont {Yoder}},\ }\href {https://doi.org/10.1103/RevModPhys.86.1337}
  {\bibfield  {journal} {\bibinfo  {journal} {Reviews of Modern Physics}\
  }\textbf {\bibinfo {volume} {86}},\ \bibinfo {pages} {1337} (\bibinfo {year}
  {2014})}\BibitemShut {NoStop}%
\bibitem [{\citenamefont {Sapra}\ \emph {et~al.}(2020)\citenamefont {Sapra},
  \citenamefont {Yang}, \citenamefont {Vercruysse}, \citenamefont {Leedle},
  \citenamefont {Black}, \citenamefont {England}, \citenamefont {Su},
  \citenamefont {Trivedi}, \citenamefont {Miao}, \citenamefont {Solgaard},
  \citenamefont {Byer},\ and\ \citenamefont {Vu{\v
  c}kovi{\'c}}}]{sapraOnchipIntegratedLaserdriven2020}%
  \BibitemOpen
  \bibfield  {author} {\bibinfo {author} {\bibfnamefont {N.~V.}\ \bibnamefont
  {Sapra}}, \bibinfo {author} {\bibfnamefont {K.~Y.}\ \bibnamefont {Yang}},
  \bibinfo {author} {\bibfnamefont {D.}~\bibnamefont {Vercruysse}}, \bibinfo
  {author} {\bibfnamefont {K.~J.}\ \bibnamefont {Leedle}}, \bibinfo {author}
  {\bibfnamefont {D.~S.}\ \bibnamefont {Black}}, \bibinfo {author}
  {\bibfnamefont {R.~J.}\ \bibnamefont {England}}, \bibinfo {author}
  {\bibfnamefont {L.}~\bibnamefont {Su}}, \bibinfo {author} {\bibfnamefont
  {R.}~\bibnamefont {Trivedi}}, \bibinfo {author} {\bibfnamefont
  {Y.}~\bibnamefont {Miao}}, \bibinfo {author} {\bibfnamefont {O.}~\bibnamefont
  {Solgaard}}, \bibinfo {author} {\bibfnamefont {R.~L.}\ \bibnamefont {Byer}},\
  and\ \bibinfo {author} {\bibfnamefont {J.}~\bibnamefont {Vu{\v
  c}kovi{\'c}}},\ }\href {https://doi.org/10.1126/science.aay5734} {\bibfield
  {journal} {\bibinfo  {journal} {Science}\ }\textbf {\bibinfo {volume}
  {367}},\ \bibinfo {pages} {79} (\bibinfo {year} {2020})}\BibitemShut
  {NoStop}%
\bibitem [{\citenamefont {Feist}\ \emph {et~al.}(2022)\citenamefont {Feist},
  \citenamefont {Huang}, \citenamefont {Arend}, \citenamefont {Yang},
  \citenamefont {Henke}, \citenamefont {Raja}, \citenamefont {Kappert},
  \citenamefont {Wang}, \citenamefont {{Louren{\c c}o-Martins}}, \citenamefont
  {Qiu}, \citenamefont {Liu}, \citenamefont {Kfir}, \citenamefont
  {Kippenberg},\ and\ \citenamefont
  {Ropers}}]{feistCavitymediatedElectronphotonPairs2022a}%
  \BibitemOpen
  \bibfield  {author} {\bibinfo {author} {\bibfnamefont {A.}~\bibnamefont
  {Feist}}, \bibinfo {author} {\bibfnamefont {G.}~\bibnamefont {Huang}},
  \bibinfo {author} {\bibfnamefont {G.}~\bibnamefont {Arend}}, \bibinfo
  {author} {\bibfnamefont {Y.}~\bibnamefont {Yang}}, \bibinfo {author}
  {\bibfnamefont {J.-W.}\ \bibnamefont {Henke}}, \bibinfo {author}
  {\bibfnamefont {A.~S.}\ \bibnamefont {Raja}}, \bibinfo {author}
  {\bibfnamefont {F.~J.}\ \bibnamefont {Kappert}}, \bibinfo {author}
  {\bibfnamefont {R.~N.}\ \bibnamefont {Wang}}, \bibinfo {author}
  {\bibfnamefont {H.}~\bibnamefont {{Louren{\c c}o-Martins}}}, \bibinfo
  {author} {\bibfnamefont {Z.}~\bibnamefont {Qiu}}, \bibinfo {author}
  {\bibfnamefont {J.}~\bibnamefont {Liu}}, \bibinfo {author} {\bibfnamefont
  {O.}~\bibnamefont {Kfir}}, \bibinfo {author} {\bibfnamefont {T.~J.}\
  \bibnamefont {Kippenberg}},\ and\ \bibinfo {author} {\bibfnamefont
  {C.}~\bibnamefont {Ropers}},\ }\href
  {https://doi.org/10.1126/science.abo5037} {\bibfield  {journal} {\bibinfo
  {journal} {Science}\ }\textbf {\bibinfo {volume} {377}},\ \bibinfo {pages}
  {777} (\bibinfo {year} {2022})}\BibitemShut {NoStop}%
\bibitem [{\citenamefont {Kippenberg}\ \emph {et~al.}(2018)\citenamefont
  {Kippenberg}, \citenamefont {Gaeta}, \citenamefont {Lipson},\ and\
  \citenamefont {Gorodetsky}}]{kippenbergDissipativeKerrSolitons2018}%
  \BibitemOpen
  \bibfield  {author} {\bibinfo {author} {\bibfnamefont {T.~J.}\ \bibnamefont
  {Kippenberg}}, \bibinfo {author} {\bibfnamefont {A.~L.}\ \bibnamefont
  {Gaeta}}, \bibinfo {author} {\bibfnamefont {M.}~\bibnamefont {Lipson}},\ and\
  \bibinfo {author} {\bibfnamefont {M.~L.}\ \bibnamefont {Gorodetsky}},\ }\href
  {https://doi.org/10.1126/science.aan8083} {\bibfield  {journal} {\bibinfo
  {journal} {Science}\ }\textbf {\bibinfo {volume} {361}},\ \bibinfo {pages}
  {eaan8083} (\bibinfo {year} {2018})}\BibitemShut {NoStop}%
\bibitem [{\citenamefont {Dudley}\ \emph {et~al.}(2006)\citenamefont {Dudley},
  \citenamefont {Genty},\ and\ \citenamefont
  {Coen}}]{dudleySupercontinuumGenerationPhotonic2006}%
  \BibitemOpen
  \bibfield  {author} {\bibinfo {author} {\bibfnamefont {J.~M.}\ \bibnamefont
  {Dudley}}, \bibinfo {author} {\bibfnamefont {G.}~\bibnamefont {Genty}},\ and\
  \bibinfo {author} {\bibfnamefont {S.}~\bibnamefont {Coen}},\ }\href
  {https://doi.org/10.1103/RevModPhys.78.1135} {\bibfield  {journal} {\bibinfo
  {journal} {Reviews of Modern Physics}\ }\textbf {\bibinfo {volume} {78}},\
  \bibinfo {pages} {1135} (\bibinfo {year} {2006})}\BibitemShut {NoStop}%
\bibitem [{\citenamefont {Grote}\ \emph {et~al.}(2013)\citenamefont {Grote},
  \citenamefont {Danzmann}, \citenamefont {Dooley}, \citenamefont {Schnabel},
  \citenamefont {Slutsky},\ and\ \citenamefont
  {Vahlbruch}}]{groteFirstLongTermApplication2013}%
  \BibitemOpen
  \bibfield  {author} {\bibinfo {author} {\bibfnamefont {H.}~\bibnamefont
  {Grote}}, \bibinfo {author} {\bibfnamefont {K.}~\bibnamefont {Danzmann}},
  \bibinfo {author} {\bibfnamefont {K.~L.}\ \bibnamefont {Dooley}}, \bibinfo
  {author} {\bibfnamefont {R.}~\bibnamefont {Schnabel}}, \bibinfo {author}
  {\bibfnamefont {J.}~\bibnamefont {Slutsky}},\ and\ \bibinfo {author}
  {\bibfnamefont {H.}~\bibnamefont {Vahlbruch}},\ }\href
  {https://doi.org/10.1103/PhysRevLett.110.181101} {\bibfield  {journal}
  {\bibinfo  {journal} {Physical Review Letters}\ }\textbf {\bibinfo {volume}
  {110}},\ \bibinfo {pages} {181101} (\bibinfo {year} {2013})}\BibitemShut
  {NoStop}%
\bibitem [{\citenamefont {Myers}\ \emph {et~al.}(1995)\citenamefont {Myers},
  \citenamefont {Eckardt}, \citenamefont {Fejer}, \citenamefont {Byer},
  \citenamefont {Bosenberg},\ and\ \citenamefont
  {Pierce}}]{myersQuasiphasematchedOpticalParametric1995}%
  \BibitemOpen
  \bibfield  {author} {\bibinfo {author} {\bibfnamefont {L.~E.}\ \bibnamefont
  {Myers}}, \bibinfo {author} {\bibfnamefont {R.~C.}\ \bibnamefont {Eckardt}},
  \bibinfo {author} {\bibfnamefont {M.~M.}\ \bibnamefont {Fejer}}, \bibinfo
  {author} {\bibfnamefont {R.~L.}\ \bibnamefont {Byer}}, \bibinfo {author}
  {\bibfnamefont {W.~R.}\ \bibnamefont {Bosenberg}},\ and\ \bibinfo {author}
  {\bibfnamefont {J.~W.}\ \bibnamefont {Pierce}},\ }\href
  {https://doi.org/10.1364/JOSAB.12.002102} {\bibfield  {journal} {\bibinfo
  {journal} {JOSA B}\ }\textbf {\bibinfo {volume} {12}},\ \bibinfo {pages}
  {2102} (\bibinfo {year} {1995})}\BibitemShut {NoStop}%
\bibitem [{\citenamefont {Kwiat}\ \emph {et~al.}(1995)\citenamefont {Kwiat},
  \citenamefont {Mattle}, \citenamefont {Weinfurter}, \citenamefont
  {Zeilinger}, \citenamefont {Sergienko},\ and\ \citenamefont
  {Shih}}]{kwiatNewHighIntensitySource1995}%
  \BibitemOpen
  \bibfield  {author} {\bibinfo {author} {\bibfnamefont {P.~G.}\ \bibnamefont
  {Kwiat}}, \bibinfo {author} {\bibfnamefont {K.}~\bibnamefont {Mattle}},
  \bibinfo {author} {\bibfnamefont {H.}~\bibnamefont {Weinfurter}}, \bibinfo
  {author} {\bibfnamefont {A.}~\bibnamefont {Zeilinger}}, \bibinfo {author}
  {\bibfnamefont {A.~V.}\ \bibnamefont {Sergienko}},\ and\ \bibinfo {author}
  {\bibfnamefont {Y.}~\bibnamefont {Shih}},\ }\href
  {https://doi.org/10.1103/PhysRevLett.75.4337} {\bibfield  {journal} {\bibinfo
   {journal} {Physical Review Letters}\ }\textbf {\bibinfo {volume} {75}},\
  \bibinfo {pages} {4337} (\bibinfo {year} {1995})}\BibitemShut {NoStop}%
\bibitem [{\citenamefont {Herr}\ \emph {et~al.}(2014)\citenamefont {Herr},
  \citenamefont {Brasch}, \citenamefont {Jost}, \citenamefont {Wang},
  \citenamefont {Kondratiev}, \citenamefont {Gorodetsky},\ and\ \citenamefont
  {Kippenberg}}]{herrTemporalSolitonsOptical2014c}%
  \BibitemOpen
  \bibfield  {author} {\bibinfo {author} {\bibfnamefont {T.}~\bibnamefont
  {Herr}}, \bibinfo {author} {\bibfnamefont {V.}~\bibnamefont {Brasch}},
  \bibinfo {author} {\bibfnamefont {J.~D.}\ \bibnamefont {Jost}}, \bibinfo
  {author} {\bibfnamefont {C.~Y.}\ \bibnamefont {Wang}}, \bibinfo {author}
  {\bibfnamefont {N.~M.}\ \bibnamefont {Kondratiev}}, \bibinfo {author}
  {\bibfnamefont {M.~L.}\ \bibnamefont {Gorodetsky}},\ and\ \bibinfo {author}
  {\bibfnamefont {T.~J.}\ \bibnamefont {Kippenberg}},\ }\href
  {https://doi.org/10.1038/nphoton.2013.343} {\bibfield  {journal} {\bibinfo
  {journal} {Nature Photonics}\ }\textbf {\bibinfo {volume} {8}},\ \bibinfo
  {pages} {145} (\bibinfo {year} {2014})}\BibitemShut {NoStop}%
\bibitem [{\citenamefont {Papp}\ \emph {et~al.}(2014)\citenamefont {Papp},
  \citenamefont {Beha}, \citenamefont {Del'Haye}, \citenamefont {Quinlan},
  \citenamefont {Lee}, \citenamefont {Vahala},\ and\ \citenamefont
  {Diddams}}]{pappMicroresonatorFrequencyComb2014}%
  \BibitemOpen
  \bibfield  {author} {\bibinfo {author} {\bibfnamefont {S.~B.}\ \bibnamefont
  {Papp}}, \bibinfo {author} {\bibfnamefont {K.}~\bibnamefont {Beha}}, \bibinfo
  {author} {\bibfnamefont {P.}~\bibnamefont {Del'Haye}}, \bibinfo {author}
  {\bibfnamefont {F.}~\bibnamefont {Quinlan}}, \bibinfo {author} {\bibfnamefont
  {H.}~\bibnamefont {Lee}}, \bibinfo {author} {\bibfnamefont {K.~J.}\
  \bibnamefont {Vahala}},\ and\ \bibinfo {author} {\bibfnamefont {S.~A.}\
  \bibnamefont {Diddams}},\ }\href {https://doi.org/10.1364/OPTICA.1.000010}
  {\bibfield  {journal} {\bibinfo  {journal} {Optica}\ }\textbf {\bibinfo
  {volume} {1}},\ \bibinfo {pages} {10} (\bibinfo {year} {2014})}\BibitemShut
  {NoStop}%
\bibitem [{\citenamefont {{Marin-Palomo}}\ \emph {et~al.}(2017)\citenamefont
  {{Marin-Palomo}}, \citenamefont {Kemal}, \citenamefont {Karpov},
  \citenamefont {Kordts}, \citenamefont {Pfeifle}, \citenamefont {Pfeiffer},
  \citenamefont {Trocha}, \citenamefont {Wolf}, \citenamefont {Brasch},
  \citenamefont {Anderson}, \citenamefont {Rosenberger}, \citenamefont
  {Vijayan}, \citenamefont {Freude}, \citenamefont {Kippenberg},\ and\
  \citenamefont {Koos}}]{marin-palomoMicroresonatorbasedSolitonsMassively2017}%
  \BibitemOpen
  \bibfield  {author} {\bibinfo {author} {\bibfnamefont {P.}~\bibnamefont
  {{Marin-Palomo}}}, \bibinfo {author} {\bibfnamefont {J.~N.}\ \bibnamefont
  {Kemal}}, \bibinfo {author} {\bibfnamefont {M.}~\bibnamefont {Karpov}},
  \bibinfo {author} {\bibfnamefont {A.}~\bibnamefont {Kordts}}, \bibinfo
  {author} {\bibfnamefont {J.}~\bibnamefont {Pfeifle}}, \bibinfo {author}
  {\bibfnamefont {M.~H.~P.}\ \bibnamefont {Pfeiffer}}, \bibinfo {author}
  {\bibfnamefont {P.}~\bibnamefont {Trocha}}, \bibinfo {author} {\bibfnamefont
  {S.}~\bibnamefont {Wolf}}, \bibinfo {author} {\bibfnamefont {V.}~\bibnamefont
  {Brasch}}, \bibinfo {author} {\bibfnamefont {M.~H.}\ \bibnamefont
  {Anderson}}, \bibinfo {author} {\bibfnamefont {R.}~\bibnamefont
  {Rosenberger}}, \bibinfo {author} {\bibfnamefont {K.}~\bibnamefont
  {Vijayan}}, \bibinfo {author} {\bibfnamefont {W.}~\bibnamefont {Freude}},
  \bibinfo {author} {\bibfnamefont {T.~J.}\ \bibnamefont {Kippenberg}},\ and\
  \bibinfo {author} {\bibfnamefont {C.}~\bibnamefont {Koos}},\ }\href
  {https://doi.org/10.1038/nature22387} {\bibfield  {journal} {\bibinfo
  {journal} {Nature}\ }\textbf {\bibinfo {volume} {546}},\ \bibinfo {pages}
  {274} (\bibinfo {year} {2017})}\BibitemShut {NoStop}%
\bibitem [{\citenamefont {Feldmann}\ \emph {et~al.}(2021)\citenamefont
  {Feldmann}, \citenamefont {Youngblood}, \citenamefont {Karpov}, \citenamefont
  {Gehring}, \citenamefont {Li}, \citenamefont {Stappers}, \citenamefont
  {Le~Gallo}, \citenamefont {Fu}, \citenamefont {Lukashchuk}, \citenamefont
  {Raja}, \citenamefont {Liu}, \citenamefont {Wright}, \citenamefont
  {Sebastian}, \citenamefont {Kippenberg}, \citenamefont {Pernice},\ and\
  \citenamefont {Bhaskaran}}]{feldmannParallelConvolutionalProcessing2021a}%
  \BibitemOpen
  \bibfield  {author} {\bibinfo {author} {\bibfnamefont {J.}~\bibnamefont
  {Feldmann}}, \bibinfo {author} {\bibfnamefont {N.}~\bibnamefont
  {Youngblood}}, \bibinfo {author} {\bibfnamefont {M.}~\bibnamefont {Karpov}},
  \bibinfo {author} {\bibfnamefont {H.}~\bibnamefont {Gehring}}, \bibinfo
  {author} {\bibfnamefont {X.}~\bibnamefont {Li}}, \bibinfo {author}
  {\bibfnamefont {M.}~\bibnamefont {Stappers}}, \bibinfo {author}
  {\bibfnamefont {M.}~\bibnamefont {Le~Gallo}}, \bibinfo {author}
  {\bibfnamefont {X.}~\bibnamefont {Fu}}, \bibinfo {author} {\bibfnamefont
  {A.}~\bibnamefont {Lukashchuk}}, \bibinfo {author} {\bibfnamefont {A.~S.}\
  \bibnamefont {Raja}}, \bibinfo {author} {\bibfnamefont {J.}~\bibnamefont
  {Liu}}, \bibinfo {author} {\bibfnamefont {C.~D.}\ \bibnamefont {Wright}},
  \bibinfo {author} {\bibfnamefont {A.}~\bibnamefont {Sebastian}}, \bibinfo
  {author} {\bibfnamefont {T.~J.}\ \bibnamefont {Kippenberg}}, \bibinfo
  {author} {\bibfnamefont {W.~H.~P.}\ \bibnamefont {Pernice}},\ and\ \bibinfo
  {author} {\bibfnamefont {H.}~\bibnamefont {Bhaskaran}},\ }\href
  {https://doi.org/10.1038/s41586-020-03070-1} {\bibfield  {journal} {\bibinfo
  {journal} {Nature}\ }\textbf {\bibinfo {volume} {589}},\ \bibinfo {pages}
  {52} (\bibinfo {year} {2021})}\BibitemShut {NoStop}%
\bibitem [{\citenamefont {Obrzud}\ \emph {et~al.}(2019)\citenamefont {Obrzud},
  \citenamefont {Rainer}, \citenamefont {Harutyunyan}, \citenamefont
  {Anderson}, \citenamefont {Liu}, \citenamefont {Geiselmann}, \citenamefont
  {Chazelas}, \citenamefont {Kundermann}, \citenamefont {Lecomte},
  \citenamefont {Cecconi}, \citenamefont {Ghedina}, \citenamefont {Molinari},
  \citenamefont {Pepe}, \citenamefont {Wildi}, \citenamefont {Bouchy},
  \citenamefont {Kippenberg},\ and\ \citenamefont
  {Herr}}]{obrzudMicrophotonicAstrocomb2019a}%
  \BibitemOpen
  \bibfield  {author} {\bibinfo {author} {\bibfnamefont {E.}~\bibnamefont
  {Obrzud}}, \bibinfo {author} {\bibfnamefont {M.}~\bibnamefont {Rainer}},
  \bibinfo {author} {\bibfnamefont {A.}~\bibnamefont {Harutyunyan}}, \bibinfo
  {author} {\bibfnamefont {M.~H.}\ \bibnamefont {Anderson}}, \bibinfo {author}
  {\bibfnamefont {J.}~\bibnamefont {Liu}}, \bibinfo {author} {\bibfnamefont
  {M.}~\bibnamefont {Geiselmann}}, \bibinfo {author} {\bibfnamefont
  {B.}~\bibnamefont {Chazelas}}, \bibinfo {author} {\bibfnamefont
  {S.}~\bibnamefont {Kundermann}}, \bibinfo {author} {\bibfnamefont
  {S.}~\bibnamefont {Lecomte}}, \bibinfo {author} {\bibfnamefont
  {M.}~\bibnamefont {Cecconi}}, \bibinfo {author} {\bibfnamefont
  {A.}~\bibnamefont {Ghedina}}, \bibinfo {author} {\bibfnamefont
  {E.}~\bibnamefont {Molinari}}, \bibinfo {author} {\bibfnamefont
  {F.}~\bibnamefont {Pepe}}, \bibinfo {author} {\bibfnamefont {F.}~\bibnamefont
  {Wildi}}, \bibinfo {author} {\bibfnamefont {F.}~\bibnamefont {Bouchy}},
  \bibinfo {author} {\bibfnamefont {T.~J.}\ \bibnamefont {Kippenberg}},\ and\
  \bibinfo {author} {\bibfnamefont {T.}~\bibnamefont {Herr}},\ }\href
  {https://doi.org/10.1038/s41566-018-0309-y} {\bibfield  {journal} {\bibinfo
  {journal} {Nature Photonics}\ }\textbf {\bibinfo {volume} {13}},\ \bibinfo
  {pages} {31} (\bibinfo {year} {2019})}\BibitemShut {NoStop}%
\bibitem [{\citenamefont {Suh}\ \emph {et~al.}(2019)\citenamefont {Suh},
  \citenamefont {Yi}, \citenamefont {Lai}, \citenamefont {Leifer},
  \citenamefont {Grudinin}, \citenamefont {Vasisht}, \citenamefont {Martin},
  \citenamefont {Fitzgerald}, \citenamefont {Doppmann}, \citenamefont {Wang},
  \citenamefont {Mawet}, \citenamefont {Papp}, \citenamefont {Diddams},
  \citenamefont {Beichman},\ and\ \citenamefont
  {Vahala}}]{suhSearchingExoplanetsUsing2019}%
  \BibitemOpen
  \bibfield  {author} {\bibinfo {author} {\bibfnamefont {M.-G.}\ \bibnamefont
  {Suh}}, \bibinfo {author} {\bibfnamefont {X.}~\bibnamefont {Yi}}, \bibinfo
  {author} {\bibfnamefont {Y.-H.}\ \bibnamefont {Lai}}, \bibinfo {author}
  {\bibfnamefont {S.}~\bibnamefont {Leifer}}, \bibinfo {author} {\bibfnamefont
  {I.~S.}\ \bibnamefont {Grudinin}}, \bibinfo {author} {\bibfnamefont
  {G.}~\bibnamefont {Vasisht}}, \bibinfo {author} {\bibfnamefont {E.~C.}\
  \bibnamefont {Martin}}, \bibinfo {author} {\bibfnamefont {M.~P.}\
  \bibnamefont {Fitzgerald}}, \bibinfo {author} {\bibfnamefont
  {G.}~\bibnamefont {Doppmann}}, \bibinfo {author} {\bibfnamefont
  {J.}~\bibnamefont {Wang}}, \bibinfo {author} {\bibfnamefont {D.}~\bibnamefont
  {Mawet}}, \bibinfo {author} {\bibfnamefont {S.~B.}\ \bibnamefont {Papp}},
  \bibinfo {author} {\bibfnamefont {S.~A.}\ \bibnamefont {Diddams}}, \bibinfo
  {author} {\bibfnamefont {C.}~\bibnamefont {Beichman}},\ and\ \bibinfo
  {author} {\bibfnamefont {K.}~\bibnamefont {Vahala}},\ }\href
  {https://doi.org/10.1038/s41566-018-0312-3} {\bibfield  {journal} {\bibinfo
  {journal} {Nature Photonics}\ }\textbf {\bibinfo {volume} {13}},\ \bibinfo
  {pages} {25} (\bibinfo {year} {2019})}\BibitemShut {NoStop}%
\bibitem [{\citenamefont {Kone{\v c}n{\'a}}\ \emph {et~al.}(2020)\citenamefont
  {Kone{\v c}n{\'a}}, \citenamefont {Di~Giulio}, \citenamefont {Mkhitaryan},
  \citenamefont {Ropers},\ and\ \citenamefont {{Garc{\'i}a de
  Abajo}}}]{konecnaNanoscaleNonlinearSpectroscopy2020b}%
  \BibitemOpen
  \bibfield  {author} {\bibinfo {author} {\bibfnamefont {A.}~\bibnamefont
  {Kone{\v c}n{\'a}}}, \bibinfo {author} {\bibfnamefont {V.}~\bibnamefont
  {Di~Giulio}}, \bibinfo {author} {\bibfnamefont {V.}~\bibnamefont
  {Mkhitaryan}}, \bibinfo {author} {\bibfnamefont {C.}~\bibnamefont {Ropers}},\
  and\ \bibinfo {author} {\bibfnamefont {F.~J.}\ \bibnamefont {{Garc{\'i}a de
  Abajo}}},\ }\href {https://doi.org/10.1021/acsphotonics.0c00326} {\bibfield
  {journal} {\bibinfo  {journal} {ACS Photonics}\ }\textbf {\bibinfo {volume}
  {7}},\ \bibinfo {pages} {1290} (\bibinfo {year} {2020})}\BibitemShut
  {NoStop}%
\bibitem [{\citenamefont {Cox}\ and\ \citenamefont {{Garc{\'i}a de
  Abajo}}(2020)}]{coxNonlinearInteractionsFree2020}%
  \BibitemOpen
  \bibfield  {author} {\bibinfo {author} {\bibfnamefont {J.~D.}\ \bibnamefont
  {Cox}}\ and\ \bibinfo {author} {\bibfnamefont {F.~J.}\ \bibnamefont
  {{Garc{\'i}a de Abajo}}},\ }\href
  {https://doi.org/10.1021/acs.nanolett.0c00538} {\bibfield  {journal}
  {\bibinfo  {journal} {Nano Letters}\ }\textbf {\bibinfo {volume} {20}},\
  \bibinfo {pages} {4792} (\bibinfo {year} {2020})}\BibitemShut {NoStop}%
\bibitem [{\citenamefont {{Garc{\'i}a de Abajo}}\ \emph
  {et~al.}(2022)\citenamefont {{Garc{\'i}a de Abajo}}, \citenamefont {Dias},\
  and\ \citenamefont
  {Di~Giulio}}]{garciadeabajoCompleteExcitationDiscrete2022}%
  \BibitemOpen
  \bibfield  {author} {\bibinfo {author} {\bibfnamefont {F.~J.}\ \bibnamefont
  {{Garc{\'i}a de Abajo}}}, \bibinfo {author} {\bibfnamefont {E.~J.~C.}\
  \bibnamefont {Dias}},\ and\ \bibinfo {author} {\bibfnamefont
  {V.}~\bibnamefont {Di~Giulio}},\ }\href
  {https://doi.org/10.1103/PhysRevLett.129.093401} {\bibfield  {journal}
  {\bibinfo  {journal} {Physical Review Letters}\ }\textbf {\bibinfo {volume}
  {129}},\ \bibinfo {pages} {093401} (\bibinfo {year} {2022})}\BibitemShut
  {NoStop}%
\bibitem [{\citenamefont {Huang}\ \emph {et~al.}(2017)\citenamefont {Huang},
  \citenamefont {Yang}, \citenamefont {Yang}, \citenamefont {Yu}, \citenamefont
  {Kwong}, \citenamefont {Zelevinsky}, \citenamefont {Jarrahi},\ and\
  \citenamefont {Wong}}]{huangGloballyStableMicroresonator2017}%
  \BibitemOpen
  \bibfield  {author} {\bibinfo {author} {\bibfnamefont {S.-W.}\ \bibnamefont
  {Huang}}, \bibinfo {author} {\bibfnamefont {J.}~\bibnamefont {Yang}},
  \bibinfo {author} {\bibfnamefont {S.-H.}\ \bibnamefont {Yang}}, \bibinfo
  {author} {\bibfnamefont {M.}~\bibnamefont {Yu}}, \bibinfo {author}
  {\bibfnamefont {D.-L.}\ \bibnamefont {Kwong}}, \bibinfo {author}
  {\bibfnamefont {T.}~\bibnamefont {Zelevinsky}}, \bibinfo {author}
  {\bibfnamefont {M.}~\bibnamefont {Jarrahi}},\ and\ \bibinfo {author}
  {\bibfnamefont {C.~W.}\ \bibnamefont {Wong}},\ }\href
  {https://doi.org/10.1103/PhysRevX.7.041002} {\bibfield  {journal} {\bibinfo
  {journal} {Physical Review X}\ }\textbf {\bibinfo {volume} {7}},\ \bibinfo
  {pages} {041002} (\bibinfo {year} {2017})}\BibitemShut {NoStop}%
\bibitem [{\citenamefont {Del'Haye}\ \emph {et~al.}(2007)\citenamefont
  {Del'Haye}, \citenamefont {Schliesser}, \citenamefont {Arcizet},
  \citenamefont {Wilken}, \citenamefont {Holzwarth},\ and\ \citenamefont
  {Kippenberg}}]{delhayeOpticalFrequencyComb2007a}%
  \BibitemOpen
  \bibfield  {author} {\bibinfo {author} {\bibfnamefont {P.}~\bibnamefont
  {Del'Haye}}, \bibinfo {author} {\bibfnamefont {A.}~\bibnamefont
  {Schliesser}}, \bibinfo {author} {\bibfnamefont {O.}~\bibnamefont {Arcizet}},
  \bibinfo {author} {\bibfnamefont {T.}~\bibnamefont {Wilken}}, \bibinfo
  {author} {\bibfnamefont {R.}~\bibnamefont {Holzwarth}},\ and\ \bibinfo
  {author} {\bibfnamefont {T.~J.}\ \bibnamefont {Kippenberg}},\ }\href
  {https://doi.org/10.1038/nature06401} {\bibfield  {journal} {\bibinfo
  {journal} {Nature}\ }\textbf {\bibinfo {volume} {450}},\ \bibinfo {pages}
  {1214} (\bibinfo {year} {2007})}\BibitemShut {NoStop}%
\bibitem [{\citenamefont {Hansson}\ \emph {et~al.}(2013)\citenamefont
  {Hansson}, \citenamefont {Modotto},\ and\ \citenamefont
  {Wabnitz}}]{hanssonDynamicsModulationalInstability2013}%
  \BibitemOpen
  \bibfield  {author} {\bibinfo {author} {\bibfnamefont {T.}~\bibnamefont
  {Hansson}}, \bibinfo {author} {\bibfnamefont {D.}~\bibnamefont {Modotto}},\
  and\ \bibinfo {author} {\bibfnamefont {S.}~\bibnamefont {Wabnitz}},\ }\href
  {https://doi.org/10.1103/PhysRevA.88.023819} {\bibfield  {journal} {\bibinfo
  {journal} {Physical Review A}\ }\textbf {\bibinfo {volume} {88}},\ \bibinfo
  {pages} {023819} (\bibinfo {year} {2013})}\BibitemShut {NoStop}%
\bibitem [{\citenamefont {Yu}\ \emph {et~al.}(2017)\citenamefont {Yu},
  \citenamefont {Jang}, \citenamefont {Okawachi}, \citenamefont {Griffith},
  \citenamefont {Luke}, \citenamefont {Miller}, \citenamefont {Ji},
  \citenamefont {Lipson},\ and\ \citenamefont
  {Gaeta}}]{yuBreatherSolitonDynamics2017}%
  \BibitemOpen
  \bibfield  {author} {\bibinfo {author} {\bibfnamefont {M.}~\bibnamefont
  {Yu}}, \bibinfo {author} {\bibfnamefont {J.~K.}\ \bibnamefont {Jang}},
  \bibinfo {author} {\bibfnamefont {Y.}~\bibnamefont {Okawachi}}, \bibinfo
  {author} {\bibfnamefont {A.~G.}\ \bibnamefont {Griffith}}, \bibinfo {author}
  {\bibfnamefont {K.}~\bibnamefont {Luke}}, \bibinfo {author} {\bibfnamefont
  {S.~A.}\ \bibnamefont {Miller}}, \bibinfo {author} {\bibfnamefont
  {X.}~\bibnamefont {Ji}}, \bibinfo {author} {\bibfnamefont {M.}~\bibnamefont
  {Lipson}},\ and\ \bibinfo {author} {\bibfnamefont {A.~L.}\ \bibnamefont
  {Gaeta}},\ }\href {https://doi.org/10.1038/ncomms14569} {\bibfield  {journal}
  {\bibinfo  {journal} {Nature Communications}\ }\textbf {\bibinfo {volume}
  {8}},\ \bibinfo {pages} {14569} (\bibinfo {year} {2017})}\BibitemShut
  {NoStop}%
\bibitem [{\citenamefont {Lucas}\ \emph {et~al.}(2017)\citenamefont {Lucas},
  \citenamefont {Karpov}, \citenamefont {Guo}, \citenamefont {Gorodetsky},\
  and\ \citenamefont {Kippenberg}}]{lucasBreathingDissipativeSolitons2017a}%
  \BibitemOpen
  \bibfield  {author} {\bibinfo {author} {\bibfnamefont {E.}~\bibnamefont
  {Lucas}}, \bibinfo {author} {\bibfnamefont {M.}~\bibnamefont {Karpov}},
  \bibinfo {author} {\bibfnamefont {H.}~\bibnamefont {Guo}}, \bibinfo {author}
  {\bibfnamefont {M.~L.}\ \bibnamefont {Gorodetsky}},\ and\ \bibinfo {author}
  {\bibfnamefont {T.~J.}\ \bibnamefont {Kippenberg}},\ }\href
  {https://doi.org/10.1038/s41467-017-00719-w} {\bibfield  {journal} {\bibinfo
  {journal} {Nature Communications}\ }\textbf {\bibinfo {volume} {8}},\
  \bibinfo {pages} {736} (\bibinfo {year} {2017})}\BibitemShut {NoStop}%
\bibitem [{\citenamefont {Cole}\ \emph {et~al.}(2017)\citenamefont {Cole},
  \citenamefont {Lamb}, \citenamefont {Del'Haye}, \citenamefont {Diddams},\
  and\ \citenamefont {Papp}}]{coleSolitonCrystalsKerr2017a}%
  \BibitemOpen
  \bibfield  {author} {\bibinfo {author} {\bibfnamefont {D.~C.}\ \bibnamefont
  {Cole}}, \bibinfo {author} {\bibfnamefont {E.~S.}\ \bibnamefont {Lamb}},
  \bibinfo {author} {\bibfnamefont {P.}~\bibnamefont {Del'Haye}}, \bibinfo
  {author} {\bibfnamefont {S.~A.}\ \bibnamefont {Diddams}},\ and\ \bibinfo
  {author} {\bibfnamefont {S.~B.}\ \bibnamefont {Papp}},\ }\href
  {https://doi.org/10.1038/s41566-017-0009-z} {\bibfield  {journal} {\bibinfo
  {journal} {Nature Photonics}\ }\textbf {\bibinfo {volume} {11}},\ \bibinfo
  {pages} {671} (\bibinfo {year} {2017})}\BibitemShut {NoStop}%
\bibitem [{\citenamefont {Karpov}\ \emph {et~al.}(2019)\citenamefont {Karpov},
  \citenamefont {Pfeiffer}, \citenamefont {Guo}, \citenamefont {Weng},
  \citenamefont {Liu},\ and\ \citenamefont
  {Kippenberg}}]{karpovDynamicsSolitonCrystals2019b}%
  \BibitemOpen
  \bibfield  {author} {\bibinfo {author} {\bibfnamefont {M.}~\bibnamefont
  {Karpov}}, \bibinfo {author} {\bibfnamefont {M.~H.~P.}\ \bibnamefont
  {Pfeiffer}}, \bibinfo {author} {\bibfnamefont {H.}~\bibnamefont {Guo}},
  \bibinfo {author} {\bibfnamefont {W.}~\bibnamefont {Weng}}, \bibinfo {author}
  {\bibfnamefont {J.}~\bibnamefont {Liu}},\ and\ \bibinfo {author}
  {\bibfnamefont {T.~J.}\ \bibnamefont {Kippenberg}},\ }\href
  {https://doi.org/10.1038/s41567-019-0635-0} {\bibfield  {journal} {\bibinfo
  {journal} {Nature Physics}\ }\textbf {\bibinfo {volume} {15}},\ \bibinfo
  {pages} {1071} (\bibinfo {year} {2019})}\BibitemShut {NoStop}%
\bibitem [{\citenamefont {Anderson}\ \emph {et~al.}(2022)\citenamefont
  {Anderson}, \citenamefont {Weng}, \citenamefont {Lihachev}, \citenamefont
  {Tikan}, \citenamefont {Liu},\ and\ \citenamefont
  {Kippenberg}}]{andersonZeroDispersionKerr2022}%
  \BibitemOpen
  \bibfield  {author} {\bibinfo {author} {\bibfnamefont {M.~H.}\ \bibnamefont
  {Anderson}}, \bibinfo {author} {\bibfnamefont {W.}~\bibnamefont {Weng}},
  \bibinfo {author} {\bibfnamefont {G.}~\bibnamefont {Lihachev}}, \bibinfo
  {author} {\bibfnamefont {A.}~\bibnamefont {Tikan}}, \bibinfo {author}
  {\bibfnamefont {J.}~\bibnamefont {Liu}},\ and\ \bibinfo {author}
  {\bibfnamefont {T.~J.}\ \bibnamefont {Kippenberg}},\ }\href
  {https://doi.org/10.1038/s41467-022-31916-x} {\bibfield  {journal} {\bibinfo
  {journal} {Nature Communications}\ }\textbf {\bibinfo {volume} {13}},\
  \bibinfo {pages} {4764} (\bibinfo {year} {2022})}\BibitemShut {NoStop}%
\bibitem [{\citenamefont {Kippenberg}\ \emph {et~al.}(2004)\citenamefont
  {Kippenberg}, \citenamefont {Spillane},\ and\ \citenamefont
  {Vahala}}]{kippenbergKerrNonlinearityOpticalParametric2004}%
  \BibitemOpen
  \bibfield  {author} {\bibinfo {author} {\bibfnamefont {T.~J.}\ \bibnamefont
  {Kippenberg}}, \bibinfo {author} {\bibfnamefont {S.~M.}\ \bibnamefont
  {Spillane}},\ and\ \bibinfo {author} {\bibfnamefont {K.~J.}\ \bibnamefont
  {Vahala}},\ }\href {https://doi.org/10.1103/PhysRevLett.93.083904} {\bibfield
   {journal} {\bibinfo  {journal} {Physical Review Letters}\ }\textbf {\bibinfo
  {volume} {93}},\ \bibinfo {pages} {083904} (\bibinfo {year}
  {2004})}\BibitemShut {NoStop}%
\bibitem [{\citenamefont {Park}\ \emph {et~al.}(2010)\citenamefont {Park},
  \citenamefont {Lin},\ and\ \citenamefont
  {Zewail}}]{parkPhotoninducedNearfieldElectron2010}%
  \BibitemOpen
  \bibfield  {author} {\bibinfo {author} {\bibfnamefont {S.~T.}\ \bibnamefont
  {Park}}, \bibinfo {author} {\bibfnamefont {M.}~\bibnamefont {Lin}},\ and\
  \bibinfo {author} {\bibfnamefont {A.~H.}\ \bibnamefont {Zewail}},\ }\href
  {https://doi.org/10.1088/1367-2630/12/12/123028} {\bibfield  {journal}
  {\bibinfo  {journal} {New Journal of Physics}\ }\textbf {\bibinfo {volume}
  {12}},\ \bibinfo {pages} {123028} (\bibinfo {year} {2010})}\BibitemShut
  {NoStop}%
\bibitem [{\citenamefont {Giulio}\ \emph {et~al.}(2019)\citenamefont {Giulio},
  \citenamefont {Kociak},\ and\ \citenamefont
  {de~Abajo}}]{giulioProbingQuantumOptical2019a}%
  \BibitemOpen
  \bibfield  {author} {\bibinfo {author} {\bibfnamefont {V.~D.}\ \bibnamefont
  {Giulio}}, \bibinfo {author} {\bibfnamefont {M.}~\bibnamefont {Kociak}},\
  and\ \bibinfo {author} {\bibfnamefont {F.~J.~G.}\ \bibnamefont {de~Abajo}},\
  }\href {https://doi.org/10.1364/OPTICA.6.001524} {\bibfield  {journal}
  {\bibinfo  {journal} {Optica}\ }\textbf {\bibinfo {volume} {6}},\ \bibinfo
  {pages} {1524} (\bibinfo {year} {2019})}\BibitemShut {NoStop}%
\bibitem [{\citenamefont {Echternkamp}\ \emph {et~al.}(2016)\citenamefont
  {Echternkamp}, \citenamefont {Feist}, \citenamefont {Sch{\"a}fer},\ and\
  \citenamefont {Ropers}}]{echternkampRamseytypePhaseControl2016a}%
  \BibitemOpen
  \bibfield  {author} {\bibinfo {author} {\bibfnamefont {K.~E.}\ \bibnamefont
  {Echternkamp}}, \bibinfo {author} {\bibfnamefont {A.}~\bibnamefont {Feist}},
  \bibinfo {author} {\bibfnamefont {S.}~\bibnamefont {Sch{\"a}fer}},\ and\
  \bibinfo {author} {\bibfnamefont {C.}~\bibnamefont {Ropers}},\ }\href
  {https://doi.org/10.1038/nphys3844} {\bibfield  {journal} {\bibinfo
  {journal} {Nature Physics}\ }\textbf {\bibinfo {volume} {12}},\ \bibinfo
  {pages} {1000} (\bibinfo {year} {2016})}\BibitemShut {NoStop}%
\bibitem [{\citenamefont {Dahan}\ \emph {et~al.}()\citenamefont {Dahan},
  \citenamefont {Gorlach}, \citenamefont {Haeusler}, \citenamefont {Karnieli},
  \citenamefont {Eyal}, \citenamefont {Yousefi}, \citenamefont {Segev},
  \citenamefont {Arie}, \citenamefont {Eisenstein}, \citenamefont
  {Hommelhoff},\ and\ \citenamefont
  {Kaminer}}]{dahanImprintingQuantumStatistics}%
  \BibitemOpen
  \bibfield  {author} {\bibinfo {author} {\bibfnamefont {R.}~\bibnamefont
  {Dahan}}, \bibinfo {author} {\bibfnamefont {A.}~\bibnamefont {Gorlach}},
  \bibinfo {author} {\bibfnamefont {U.}~\bibnamefont {Haeusler}}, \bibinfo
  {author} {\bibfnamefont {A.}~\bibnamefont {Karnieli}}, \bibinfo {author}
  {\bibfnamefont {O.}~\bibnamefont {Eyal}}, \bibinfo {author} {\bibfnamefont
  {P.}~\bibnamefont {Yousefi}}, \bibinfo {author} {\bibfnamefont
  {M.}~\bibnamefont {Segev}}, \bibinfo {author} {\bibfnamefont
  {A.}~\bibnamefont {Arie}}, \bibinfo {author} {\bibfnamefont {G.}~\bibnamefont
  {Eisenstein}}, \bibinfo {author} {\bibfnamefont {P.}~\bibnamefont
  {Hommelhoff}},\ and\ \bibinfo {author} {\bibfnamefont {I.}~\bibnamefont
  {Kaminer}},\ }\href {https://doi.org/10.1126/science.abj7128} {\bibfield
  {journal} {\bibinfo  {journal} {Science}\ }\textbf {\bibinfo {volume}
  {373}},\ \bibinfo {pages} {eabj7128}}\BibitemShut {NoStop}%
\bibitem [{\citenamefont {Guo}\ \emph {et~al.}(2017)\citenamefont {Guo},
  \citenamefont {Karpov}, \citenamefont {Lucas}, \citenamefont {Kordts},
  \citenamefont {Pfeiffer}, \citenamefont {Brasch}, \citenamefont {Lihachev},
  \citenamefont {Lobanov}, \citenamefont {Gorodetsky},\ and\ \citenamefont
  {Kippenberg}}]{guoUniversalDynamicsDeterministic2017}%
  \BibitemOpen
  \bibfield  {author} {\bibinfo {author} {\bibfnamefont {H.}~\bibnamefont
  {Guo}}, \bibinfo {author} {\bibfnamefont {M.}~\bibnamefont {Karpov}},
  \bibinfo {author} {\bibfnamefont {E.}~\bibnamefont {Lucas}}, \bibinfo
  {author} {\bibfnamefont {A.}~\bibnamefont {Kordts}}, \bibinfo {author}
  {\bibfnamefont {M.~H.~P.}\ \bibnamefont {Pfeiffer}}, \bibinfo {author}
  {\bibfnamefont {V.}~\bibnamefont {Brasch}}, \bibinfo {author} {\bibfnamefont
  {G.}~\bibnamefont {Lihachev}}, \bibinfo {author} {\bibfnamefont {V.~E.}\
  \bibnamefont {Lobanov}}, \bibinfo {author} {\bibfnamefont {M.~L.}\
  \bibnamefont {Gorodetsky}},\ and\ \bibinfo {author} {\bibfnamefont {T.~J.}\
  \bibnamefont {Kippenberg}},\ }\href {https://doi.org/10.1038/nphys3893}
  {\bibfield  {journal} {\bibinfo  {journal} {Nature Physics}\ }\textbf
  {\bibinfo {volume} {13}},\ \bibinfo {pages} {94} (\bibinfo {year}
  {2017})}\BibitemShut {NoStop}%
\bibitem [{\citenamefont {Talebi}(2020)}]{talebiStrongInteractionSlow2020}%
  \BibitemOpen
  \bibfield  {author} {\bibinfo {author} {\bibfnamefont {N.}~\bibnamefont
  {Talebi}},\ }\href {https://doi.org/10.1103/PhysRevLett.125.080401}
  {\bibfield  {journal} {\bibinfo  {journal} {Physical Review Letters}\
  }\textbf {\bibinfo {volume} {125}},\ \bibinfo {pages} {080401} (\bibinfo
  {year} {2020})}\BibitemShut {NoStop}%
\bibitem [{\citenamefont {Chahshouri}\ and\ \citenamefont
  {Talebi}(2023)}]{chahshouriTailoringNearfieldmediatedPhoton2023}%
  \BibitemOpen
  \bibfield  {author} {\bibinfo {author} {\bibfnamefont {F.}~\bibnamefont
  {Chahshouri}}\ and\ \bibinfo {author} {\bibfnamefont {N.}~\bibnamefont
  {Talebi}},\ }\href {https://doi.org/10.1088/1367-2630/acb4b7} {\bibfield
  {journal} {\bibinfo  {journal} {New Journal of Physics}\ }\textbf {\bibinfo
  {volume} {25}},\ \bibinfo {pages} {013033} (\bibinfo {year}
  {2023})}\BibitemShut {NoStop}%
\bibitem [{\citenamefont {Hassan}\ \emph {et~al.}(2015)\citenamefont {Hassan},
  \citenamefont {Liu}, \citenamefont {Baskin},\ and\ \citenamefont
  {Zewail}}]{hassanPhotonGatingFourdimensional2015}%
  \BibitemOpen
  \bibfield  {author} {\bibinfo {author} {\bibfnamefont {M.~T.}\ \bibnamefont
  {Hassan}}, \bibinfo {author} {\bibfnamefont {H.}~\bibnamefont {Liu}},
  \bibinfo {author} {\bibfnamefont {J.~S.}\ \bibnamefont {Baskin}},\ and\
  \bibinfo {author} {\bibfnamefont {A.~H.}\ \bibnamefont {Zewail}},\ }\href
  {https://doi.org/10.1073/pnas.1517942112} {\bibfield  {journal} {\bibinfo
  {journal} {Proceedings of the National Academy of Sciences}\ }\textbf
  {\bibinfo {volume} {112}},\ \bibinfo {pages} {12944} (\bibinfo {year}
  {2015})}\BibitemShut {NoStop}%
\bibitem [{\citenamefont {Fu}\ \emph {et~al.}(2020)\citenamefont {Fu},
  \citenamefont {Barantani}, \citenamefont {Gargiulo}, \citenamefont {Madan},
  \citenamefont {Berruto}, \citenamefont {LaGrange}, \citenamefont {Jin},
  \citenamefont {Wu}, \citenamefont {Vanacore}, \citenamefont {Carbone},\ and\
  \citenamefont {Zhu}}]{fuNanoscalefemtosecondDielectricResponse2020}%
  \BibitemOpen
  \bibfield  {author} {\bibinfo {author} {\bibfnamefont {X.}~\bibnamefont
  {Fu}}, \bibinfo {author} {\bibfnamefont {F.}~\bibnamefont {Barantani}},
  \bibinfo {author} {\bibfnamefont {S.}~\bibnamefont {Gargiulo}}, \bibinfo
  {author} {\bibfnamefont {I.}~\bibnamefont {Madan}}, \bibinfo {author}
  {\bibfnamefont {G.}~\bibnamefont {Berruto}}, \bibinfo {author} {\bibfnamefont
  {T.}~\bibnamefont {LaGrange}}, \bibinfo {author} {\bibfnamefont
  {L.}~\bibnamefont {Jin}}, \bibinfo {author} {\bibfnamefont {J.}~\bibnamefont
  {Wu}}, \bibinfo {author} {\bibfnamefont {G.~M.}\ \bibnamefont {Vanacore}},
  \bibinfo {author} {\bibfnamefont {F.}~\bibnamefont {Carbone}},\ and\ \bibinfo
  {author} {\bibfnamefont {Y.}~\bibnamefont {Zhu}},\ }\href
  {https://doi.org/10.1038/s41467-020-19636-6} {\bibfield  {journal} {\bibinfo
  {journal} {Nature Communications}\ }\textbf {\bibinfo {volume} {11}},\
  \bibinfo {pages} {5770} (\bibinfo {year} {2020})}\BibitemShut {NoStop}%
\bibitem [{\citenamefont {Black}\ \emph {et~al.}(2019)\citenamefont {Black},
  \citenamefont {Niedermayer}, \citenamefont {Miao}, \citenamefont {Zhao},
  \citenamefont {Solgaard}, \citenamefont {Byer},\ and\ \citenamefont
  {Leedle}}]{blackNetAccelerationDirect2019a}%
  \BibitemOpen
  \bibfield  {author} {\bibinfo {author} {\bibfnamefont {D.~S.}\ \bibnamefont
  {Black}}, \bibinfo {author} {\bibfnamefont {U.}~\bibnamefont {Niedermayer}},
  \bibinfo {author} {\bibfnamefont {Y.}~\bibnamefont {Miao}}, \bibinfo {author}
  {\bibfnamefont {Z.}~\bibnamefont {Zhao}}, \bibinfo {author} {\bibfnamefont
  {O.}~\bibnamefont {Solgaard}}, \bibinfo {author} {\bibfnamefont {R.~L.}\
  \bibnamefont {Byer}},\ and\ \bibinfo {author} {\bibfnamefont {K.~J.}\
  \bibnamefont {Leedle}},\ }\href
  {https://doi.org/10.1103/PhysRevLett.123.264802} {\bibfield  {journal}
  {\bibinfo  {journal} {Physical Review Letters}\ }\textbf {\bibinfo {volume}
  {123}},\ \bibinfo {pages} {264802} (\bibinfo {year} {2019})}\BibitemShut
  {NoStop}%
\bibitem [{\citenamefont {Sch{\"o}nenberger}\ \emph {et~al.}(2019)\citenamefont
  {Sch{\"o}nenberger}, \citenamefont {Mittelbach}, \citenamefont {Yousefi},
  \citenamefont {McNeur}, \citenamefont {Niedermayer},\ and\ \citenamefont
  {Hommelhoff}}]{schonenbergerGenerationCharacterizationAttosecond2019a}%
  \BibitemOpen
  \bibfield  {author} {\bibinfo {author} {\bibfnamefont {N.}~\bibnamefont
  {Sch{\"o}nenberger}}, \bibinfo {author} {\bibfnamefont {A.}~\bibnamefont
  {Mittelbach}}, \bibinfo {author} {\bibfnamefont {P.}~\bibnamefont {Yousefi}},
  \bibinfo {author} {\bibfnamefont {J.}~\bibnamefont {McNeur}}, \bibinfo
  {author} {\bibfnamefont {U.}~\bibnamefont {Niedermayer}},\ and\ \bibinfo
  {author} {\bibfnamefont {P.}~\bibnamefont {Hommelhoff}},\ }\href
  {https://doi.org/10.1103/PhysRevLett.123.264803} {\bibfield  {journal}
  {\bibinfo  {journal} {Physical Review Letters}\ }\textbf {\bibinfo {volume}
  {123}},\ \bibinfo {pages} {264803} (\bibinfo {year} {2019})}\BibitemShut
  {NoStop}%
\bibitem [{\citenamefont {Yalunin}\ \emph {et~al.}(2021)\citenamefont
  {Yalunin}, \citenamefont {Feist},\ and\ \citenamefont
  {Ropers}}]{yaluninTailoredHighcontrastAttosecond2021}%
  \BibitemOpen
  \bibfield  {author} {\bibinfo {author} {\bibfnamefont {S.~V.}\ \bibnamefont
  {Yalunin}}, \bibinfo {author} {\bibfnamefont {A.}~\bibnamefont {Feist}},\
  and\ \bibinfo {author} {\bibfnamefont {C.}~\bibnamefont {Ropers}},\ }\href
  {https://doi.org/10.1103/PhysRevResearch.3.L032036} {\bibfield  {journal}
  {\bibinfo  {journal} {Physical Review Research}\ }\textbf {\bibinfo {volume}
  {3}},\ \bibinfo {pages} {L032036} (\bibinfo {year} {2021})}\BibitemShut
  {NoStop}%
\bibitem [{\citenamefont {Pfeiffer}\ \emph {et~al.}(2018)\citenamefont
  {Pfeiffer}, \citenamefont {Liu}, \citenamefont {Raja}, \citenamefont
  {Morais}, \citenamefont {Ghadiani},\ and\ \citenamefont
  {Kippenberg}}]{pfeifferUltrasmoothSiliconNitride2018b}%
  \BibitemOpen
  \bibfield  {author} {\bibinfo {author} {\bibfnamefont {M.~H.~P.}\
  \bibnamefont {Pfeiffer}}, \bibinfo {author} {\bibfnamefont {J.}~\bibnamefont
  {Liu}}, \bibinfo {author} {\bibfnamefont {A.~S.}\ \bibnamefont {Raja}},
  \bibinfo {author} {\bibfnamefont {T.}~\bibnamefont {Morais}}, \bibinfo
  {author} {\bibfnamefont {B.}~\bibnamefont {Ghadiani}},\ and\ \bibinfo
  {author} {\bibfnamefont {T.~J.}\ \bibnamefont {Kippenberg}},\ }\href
  {https://doi.org/10.1364/OPTICA.5.000884} {\bibfield  {journal} {\bibinfo
  {journal} {Optica}\ }\textbf {\bibinfo {volume} {5}},\ \bibinfo {pages} {884}
  (\bibinfo {year} {2018})}\BibitemShut {NoStop}%
\bibitem [{\citenamefont {Del'Haye}\ \emph {et~al.}(2009)\citenamefont
  {Del'Haye}, \citenamefont {Arcizet}, \citenamefont {Gorodetsky},
  \citenamefont {Holzwarth},\ and\ \citenamefont
  {Kippenberg}}]{delhayeFrequencyCombAssisted2009a}%
  \BibitemOpen
  \bibfield  {author} {\bibinfo {author} {\bibfnamefont {P.}~\bibnamefont
  {Del'Haye}}, \bibinfo {author} {\bibfnamefont {O.}~\bibnamefont {Arcizet}},
  \bibinfo {author} {\bibfnamefont {M.~L.}\ \bibnamefont {Gorodetsky}},
  \bibinfo {author} {\bibfnamefont {R.}~\bibnamefont {Holzwarth}},\ and\
  \bibinfo {author} {\bibfnamefont {T.~J.}\ \bibnamefont {Kippenberg}},\ }\href
  {https://doi.org/10.1038/nphoton.2009.138} {\bibfield  {journal} {\bibinfo
  {journal} {Nature Photonics}\ }\textbf {\bibinfo {volume} {3}},\ \bibinfo
  {pages} {529} (\bibinfo {year} {2009})}\BibitemShut {NoStop}%
\bibitem [{\citenamefont {Feist}\ \emph {et~al.}(2017)\citenamefont {Feist},
  \citenamefont {Bach}, \citenamefont {{Rubiano da Silva}}, \citenamefont
  {Danz}, \citenamefont {M{\"o}ller}, \citenamefont {Priebe}, \citenamefont
  {Domr{\"o}se}, \citenamefont {Gatzmann}, \citenamefont {Rost}, \citenamefont
  {Schauss}, \citenamefont {Strauch}, \citenamefont {Bormann}, \citenamefont
  {Sivis}, \citenamefont {Sch{\"a}fer},\ and\ \citenamefont
  {Ropers}}]{feistUltrafastTransmissionElectron2017}%
  \BibitemOpen
  \bibfield  {author} {\bibinfo {author} {\bibfnamefont {A.}~\bibnamefont
  {Feist}}, \bibinfo {author} {\bibfnamefont {N.}~\bibnamefont {Bach}},
  \bibinfo {author} {\bibfnamefont {N.}~\bibnamefont {{Rubiano da Silva}}},
  \bibinfo {author} {\bibfnamefont {T.}~\bibnamefont {Danz}}, \bibinfo {author}
  {\bibfnamefont {M.}~\bibnamefont {M{\"o}ller}}, \bibinfo {author}
  {\bibfnamefont {K.~E.}\ \bibnamefont {Priebe}}, \bibinfo {author}
  {\bibfnamefont {T.}~\bibnamefont {Domr{\"o}se}}, \bibinfo {author}
  {\bibfnamefont {J.~G.}\ \bibnamefont {Gatzmann}}, \bibinfo {author}
  {\bibfnamefont {S.}~\bibnamefont {Rost}}, \bibinfo {author} {\bibfnamefont
  {J.}~\bibnamefont {Schauss}}, \bibinfo {author} {\bibfnamefont
  {S.}~\bibnamefont {Strauch}}, \bibinfo {author} {\bibfnamefont
  {R.}~\bibnamefont {Bormann}}, \bibinfo {author} {\bibfnamefont
  {M.}~\bibnamefont {Sivis}}, \bibinfo {author} {\bibfnamefont
  {S.}~\bibnamefont {Sch{\"a}fer}},\ and\ \bibinfo {author} {\bibfnamefont
  {C.}~\bibnamefont {Ropers}},\ }\href
  {https://doi.org/10.1016/j.ultramic.2016.12.005} {\bibfield  {journal}
  {\bibinfo  {journal} {Ultramicroscopy}\ }\bibinfo {series} {70th {{Birthday}}
  of {{Robert Sinclair}} and 65th {{Birthday}} of {{Nestor J}}. {{Zaluzec
  PICO}} 2017 \textendash{} {{Fourth Conference}} on {{Frontiers}} of
  {{Aberration Corrected Electron Microscopy}}},\ \textbf {\bibinfo {volume}
  {176}},\ \bibinfo {pages} {63} (\bibinfo {year} {2017})}\BibitemShut
  {NoStop}%
\bibitem [{\citenamefont {Koz{\'a}k}\ \emph {et~al.}(2017)\citenamefont
  {Koz{\'a}k}, \citenamefont {Beck}, \citenamefont {Deng}, \citenamefont
  {McNeur}, \citenamefont {Sch{\"o}nenberger}, \citenamefont {Gaida},
  \citenamefont {Stutzki}, \citenamefont {Gebhardt}, \citenamefont {Limpert},
  \citenamefont {Ruehl}, \citenamefont {Hartl}, \citenamefont {Solgaard},
  \citenamefont {Harris}, \citenamefont {Byer},\ and\ \citenamefont
  {Hommelhoff}}]{kozakAccelerationSubrelativisticElectrons2017}%
  \BibitemOpen
  \bibfield  {author} {\bibinfo {author} {\bibfnamefont {M.}~\bibnamefont
  {Koz{\'a}k}}, \bibinfo {author} {\bibfnamefont {P.}~\bibnamefont {Beck}},
  \bibinfo {author} {\bibfnamefont {H.}~\bibnamefont {Deng}}, \bibinfo {author}
  {\bibfnamefont {J.}~\bibnamefont {McNeur}}, \bibinfo {author} {\bibfnamefont
  {N.}~\bibnamefont {Sch{\"o}nenberger}}, \bibinfo {author} {\bibfnamefont
  {C.}~\bibnamefont {Gaida}}, \bibinfo {author} {\bibfnamefont
  {F.}~\bibnamefont {Stutzki}}, \bibinfo {author} {\bibfnamefont
  {M.}~\bibnamefont {Gebhardt}}, \bibinfo {author} {\bibfnamefont
  {J.}~\bibnamefont {Limpert}}, \bibinfo {author} {\bibfnamefont
  {A.}~\bibnamefont {Ruehl}}, \bibinfo {author} {\bibfnamefont
  {I.}~\bibnamefont {Hartl}}, \bibinfo {author} {\bibfnamefont
  {O.}~\bibnamefont {Solgaard}}, \bibinfo {author} {\bibfnamefont {J.~S.}\
  \bibnamefont {Harris}}, \bibinfo {author} {\bibfnamefont {R.~L.}\
  \bibnamefont {Byer}},\ and\ \bibinfo {author} {\bibfnamefont
  {P.}~\bibnamefont {Hommelhoff}},\ }\href
  {https://doi.org/10.1364/OE.25.019195} {\bibfield  {journal} {\bibinfo
  {journal} {Optics Express}\ }\textbf {\bibinfo {volume} {25}},\ \bibinfo
  {pages} {19195} (\bibinfo {year} {2017})}\BibitemShut {NoStop}%
\bibitem [{\citenamefont {Dahan}\ \emph {et~al.}(2020)\citenamefont {Dahan},
  \citenamefont {Nehemia}, \citenamefont {Shentcis}, \citenamefont {Reinhardt},
  \citenamefont {Adiv}, \citenamefont {Shi}, \citenamefont {Be'er},
  \citenamefont {Lynch}, \citenamefont {Kurman}, \citenamefont {Wang},\ and\
  \citenamefont {Kaminer}}]{dahanResonantPhasematchingLight2020}%
  \BibitemOpen
  \bibfield  {author} {\bibinfo {author} {\bibfnamefont {R.}~\bibnamefont
  {Dahan}}, \bibinfo {author} {\bibfnamefont {S.}~\bibnamefont {Nehemia}},
  \bibinfo {author} {\bibfnamefont {M.}~\bibnamefont {Shentcis}}, \bibinfo
  {author} {\bibfnamefont {O.}~\bibnamefont {Reinhardt}}, \bibinfo {author}
  {\bibfnamefont {Y.}~\bibnamefont {Adiv}}, \bibinfo {author} {\bibfnamefont
  {X.}~\bibnamefont {Shi}}, \bibinfo {author} {\bibfnamefont {O.}~\bibnamefont
  {Be'er}}, \bibinfo {author} {\bibfnamefont {M.~H.}\ \bibnamefont {Lynch}},
  \bibinfo {author} {\bibfnamefont {Y.}~\bibnamefont {Kurman}}, \bibinfo
  {author} {\bibfnamefont {K.}~\bibnamefont {Wang}},\ and\ \bibinfo {author}
  {\bibfnamefont {I.}~\bibnamefont {Kaminer}},\ }\href
  {https://doi.org/10.1038/s41567-020-01042-w} {\bibfield  {journal} {\bibinfo
  {journal} {Nature Physics}\ }\textbf {\bibinfo {volume} {16}},\ \bibinfo
  {pages} {1123} (\bibinfo {year} {2020})}\BibitemShut {NoStop}%
\bibitem [{\citenamefont {Brasch}\ \emph {et~al.}(2014)\citenamefont {Brasch},
  \citenamefont {Chen}, \citenamefont {Schiller},\ and\ \citenamefont
  {Kippenberg}}]{braschRadiationHardnessHighQ2014}%
  \BibitemOpen
  \bibfield  {author} {\bibinfo {author} {\bibfnamefont {V.}~\bibnamefont
  {Brasch}}, \bibinfo {author} {\bibfnamefont {Q.-F.}\ \bibnamefont {Chen}},
  \bibinfo {author} {\bibfnamefont {S.}~\bibnamefont {Schiller}},\ and\
  \bibinfo {author} {\bibfnamefont {T.~J.}\ \bibnamefont {Kippenberg}},\ }\href
  {https://doi.org/10.1364/OE.22.030786} {\bibfield  {journal} {\bibinfo
  {journal} {Optics Express}\ }\textbf {\bibinfo {volume} {22}},\ \bibinfo
  {pages} {30786} (\bibinfo {year} {2014})}\BibitemShut {NoStop}%
\bibitem [{\citenamefont {Brasch}\ \emph {et~al.}(2016)\citenamefont {Brasch},
  \citenamefont {Geiselmann}, \citenamefont {Herr}, \citenamefont {Lihachev},
  \citenamefont {Pfeiffer}, \citenamefont {Gorodetsky},\ and\ \citenamefont
  {Kippenberg}}]{braschPhotonicChipBased2016c}%
  \BibitemOpen
  \bibfield  {author} {\bibinfo {author} {\bibfnamefont {V.}~\bibnamefont
  {Brasch}}, \bibinfo {author} {\bibfnamefont {M.}~\bibnamefont {Geiselmann}},
  \bibinfo {author} {\bibfnamefont {T.}~\bibnamefont {Herr}}, \bibinfo {author}
  {\bibfnamefont {G.}~\bibnamefont {Lihachev}}, \bibinfo {author}
  {\bibfnamefont {M.~H.~P.}\ \bibnamefont {Pfeiffer}}, \bibinfo {author}
  {\bibfnamefont {M.~L.}\ \bibnamefont {Gorodetsky}},\ and\ \bibinfo {author}
  {\bibfnamefont {T.~J.}\ \bibnamefont {Kippenberg}},\ }\href
  {https://doi.org/10.1126/science.aad4811} {\bibfield  {journal} {\bibinfo
  {journal} {Science}\ }\textbf {\bibinfo {volume} {351}},\ \bibinfo {pages}
  {357} (\bibinfo {year} {2016})}\BibitemShut {NoStop}%
\bibitem [{\citenamefont {{Garc{\'i}a de Abajo}}\ and\ \citenamefont {Kone{\v
  c}n{\'a}}(2021)}]{garciadeabajoOpticalModulationElectron2021}%
  \BibitemOpen
  \bibfield  {author} {\bibinfo {author} {\bibfnamefont {F.~J.}\ \bibnamefont
  {{Garc{\'i}a de Abajo}}}\ and\ \bibinfo {author} {\bibfnamefont
  {A.}~\bibnamefont {Kone{\v c}n{\'a}}},\ }\href
  {https://doi.org/10.1103/PhysRevLett.126.123901} {\bibfield  {journal}
  {\bibinfo  {journal} {Physical Review Letters}\ }\textbf {\bibinfo {volume}
  {126}},\ \bibinfo {pages} {123901} (\bibinfo {year} {2021})}\BibitemShut
  {NoStop}%
\end{thebibliography}%
\end{document}